\documentclass[aps,pra,twocolumn,superscriptaddress]{revtex4-1}
\usepackage{amsmath}
\usepackage{amssymb}
\usepackage{mathrsfs}
\usepackage{graphicx}
\usepackage{siunitx}
\usepackage{enumerate}

\usepackage{multirow}
\usepackage{mathrsfs}

\newcommand{\ve}[1]{\mathbf{#1}}

\def\bra#1{\mathinner{\langle{#1}|}}
\def\ket#1{\mathinner{|{#1}\rangle}}

\def\Bra#1{\left<#1\right|}

{\catcode`\|=\active\gdef\Bra5t5yket#1{\left<\mathcode`\|"8000\let|\bravert {#1}\right>}}
\def\bravert{\egroup\,\vrule\,\bgroup}

\newcommand{\vac}{\ket{\text{vac}}}

\begin{document}


\title{Third order parametric downconversion: a stimulated approach}


\author{Francisco A. Dom\'inguez-Serna}
\email[]{fadomin@cicese.mx}
\affiliation{C\'atedras CONACYT, Centro de Investigaci\'on Cient\'fica y de Educaci\'on Superior de Ensenada, Apartado Postal 2732, BC 22860 Ensenada, M\'exico}
\affiliation{Departamento de \'Optica, Centro de Investigaci\'on Cient\'ifica y de Educaci\'on Superior de Ensenada, Apartado Postal 2732, BC 22860 Ensenada, M\'exico}

\author{Alfred B. U'Ren}
\email[]{alfred.uren@correo.nucleares.unam.mx}
\affiliation{Instituto de Ciencias Nucleares, Universidad Nacional Aut\'onoma de M\'exico, apdo. postal 70-543,04510 Ciudad de M\'exico, M\'exico}

\author{Karina Garay-Palmett}
\email[]{kgaray@cicese.mx}
\affiliation{Departamento de \'Optica, Centro de Investigaci\'on Cient\'ifica y de Educaci\'on Superior de Ensenada, Apartado Postal 2732, BC 22860 Ensenada, M\'exico}


\date{\today}

\begin{abstract}
We study the process of seeded, or stimulated,  third-order parametric down-conversion, as an extension of our previous work on spontaneous parametric downconversion (TOSPDC).   We present general expressions for the spectra and throughputs expected for the cases where the seed field or fields overlap either only one or two of the TOSPDC modes, and also allow for both pump and seed to be either monochromatic or pulsed.   We present a numerical study for a particular source design, showing that doubly-overlapped seeding can lead to a considerably greater generated flux as compared with singly-overlapped seeding.  We furthermore show that doubly-overlapped seeding permits stimulated emission tomography for the reconstruction of the three-photon TOSPDC joint spectral intensity.   We hope that our work will guide future experimental efforts based on the process of third-order  parametric downconversion.
\end{abstract}


\maketitle

\section{Introduction \label{sec:intro}}

The promise of quantum-enabled technologies, which can outperform their counterparts based on classical physics, has motivated a number of exciting lines of research \cite{Bennett2000,Jozsa2003,Hughes2007,Prevedel2007,Vedral2014,Brecht2015}. While a definitive technology for the implementation of all quantum information science (QIS) tasks does not exist, it is in general believed that photons are well suited for some of these tasks  \cite{Obrien2010}.  Such photonics-based QIS leads to the need for sources of single-photons \cite{Knill01a} and  of multiple photons in quantum-entangled states  \cite{OBrien2007,Vedral2014}.  In this paper, we present a study of seeded third-order parametric downconversion as a route towards the characterization and utilization of three-photon states.
 
Nowadays, the use of nonlinear spontaneous parametric processes for the generation of entangled photon pairs and heralded single photons has become standard  \cite{Burnham1970,Cohen2009,Pittman2005,Huang2011,Meyer-Scott2017,Zhang2012}. However, the generation of \emph{heralded photon pairs} that requires the availability of photon triplet sources, remains challenging.   While cascaded sources of photon triplets, i.e. relying on an initial photon pair generation stage with one of the generated modes later acting as pump for a second photon pair generation stage, have been demonstrated \cite{Sliwa2003,Walther2007,Barz2010, Niu2009,Hubel2010,Hamel2014,Agne2017,Krapick2016}, the development of genuine photon triplet sources in which the three photons derive from a single quantum event is an ongoing research topic \cite{Hamel2014}.

One possible avenue towards the above goal is the use of the process of third-order spontaneous  parametric downconversion (TOSPDC), in which a pump photon is annihilated with the consequent creation of a photon triplet, in such a manner that energy and momentum are conserved.  TOSPDC is a direct generalization, relying on a $\chi^{(3)}$ nonlinearity, of the well known second-order spontaneonous parametric downconversion process (SPDC), mediated by a $\chi^{(2)}$ nonlinearity.  TOSPDC has been explored theoretically, at first in a hypothetical medium with third order non-linearity \cite{Chekhova2005}, and later through specific proposals:  one, from our group, relying on the use of a  thin cylindrical waveguide surrounded by air, which could be realized in the form of a tapered fiber \cite{Corona2011,Corona2011a}; and another based on nonlinear crystals \cite{Dot2012}. Note that given the very large spectral separation between the pump (at $\omega_p$)  and the generated photons (around $\omega_p/3$) inherent in TOSPDC,  phasematching involving all four waves in the fundamental fiber mode is, for fused silica and other fiber materials, not feasible. Thus, our proposal relies on intra-modal phasematching with the pump in a non-fundamental mode and the photon triplets in the fundamental mode.   While phasematching can indeed be attained in this manner,  the challenge now becomes the fact that the emission rate, which is proportional to the overlap integral between the four interacting waves is very low for such intra-modal phasematching.  Under ideal conditions and  with realistic experimental parameters, the expected emission rate from such a source is $<10$ triplets/s.  

As with photon pair generation, the characterization of spectral emission properties, including spectral entanglement, would become a key aspect of photon triplet experiments. A comprehensive review of spectral characterization techniques for photon pairs can be found in Ref. \cite{Zielnicki2018}.     Such  spectral characterization of photon pairs can be time consuming, particularly if rasterization techniques are used.   Given the very low conversion efficiency of TOSPDC, such rasterization techniques would in all likelihood be unfeasible.     

Among the techniques reviewed in \cite{Zielnicki2018}, stimulated emission tomography (SET) based on a stimulated version of the parametric process is promising. As first explored in \cite{Liscidini2013} for photon pair sources, it is possible to utilize a tunable seed, say at the idler frequency $\omega_i$, so as to stimulate emission at the signal mode, which can then be measured with the help of a standard spectrometer for classical light.   The authors demonstrated that the rate of spontaneous generation in the idler mode can be inferred as the quotient of the stimulated emission rate for the signal mode to the incoming seed  power in the idler mode.  By combining the signal-mode (classical) spectra obtained for the various idler frequencies it then becomes possible to obtain a 2D frequency map that corresponds to the joint spectral intensity, which would have been obtained through a quantum measurement of the unseeded, spontaneous source.  The theoretical proposal by Liscidini et al. was later implemented experimentally \cite{Fang2016,Fang2014}.  One of the motivations behind the present work is to explore whether this idea can be extended to photon-triplet sources.  In this regard, the pioneering work of Dot \textit{et.al.} \cite{Dot2012} described theoretically spontaneous and stimulated generation in the third order parametric downconversion process in the possible presence of a seed or seeds, with experimental work by the same authors reported in \cite{Gravier2008,Douady2004}. In addition,  Okoth \textit{et.al.}  \cite{Okoth2018} have analyzed theoretically  seeding for third and higher-order parametric processes.

In this work, we present the theory for the process of stimulated third-order parametric downconversion process, as a generalization of the scheme proposed by  Liscidini et al. in which we employ a fully quantum description of all the fields involved.    Note that henceforth we use the abbreviation STOPDC for stimulated third-order parametric downconversion, in contrast to TOSPDC for third-order spontaneous parametric downconversion.  As part of this description we study different configurations for the pump and seed fields, and  derive expressions for the resulting output flux for each of the fields involved, as an important guide for future experiments.   In addition, we  explore the possibility of exploiting the STOPDC process as the basis for the spectral  characterization of the three-photon state, in analogy to the stimulated emission tomography technique already demonstrated for photon-pair sources.  We hope that the present work will pave the road towards the full exploitation of the process of third order parametric downconversion, particularly in seeded configurations.

\section{Quantum state produced by STOPDC}

\begin{figure}
	\includegraphics[width=0.48\textwidth]{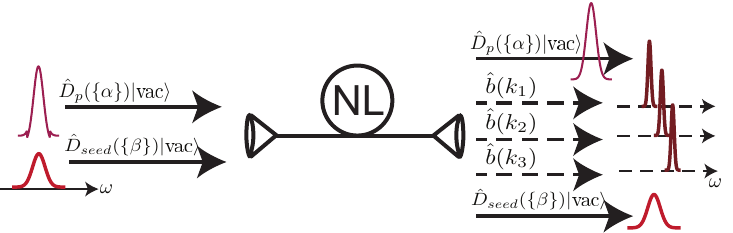}
	\caption{Schematic for the process of  stimulated, or seeded, third-order parametric downconversion. NL represents the third order nonlinear material. Pump and seed fields are described as coherent states $\hat{D}_p (\{ \alpha\})$, $\hat{D}_{seed} (\{ \beta\}) \vac$. $\hat{b}(k_1)$, $\hat{b}(k_2)$, and $\hat{b}(k_3)$ are the generation modes. }
	\label{Fig:CTN}
\end{figure}

Figure \ref{Fig:CTN} depicts the STOPDC process schematically.  NL represents the non-linear material with a $\chi^{(3)}$ nonlinearity, assumed to be in the form of an optical fiber.  The pump field is described by the coherent field $\hat{D}_p (\{ \alpha\}) \vac$, in terms of the displacement operator $\hat{D}_p (\{ \alpha\})$. The operators $\hat{b}(k_i)$ with $i=\{ 1,2,3\}$ correspond to the energy-conserving and phase-matched generation modes of the spontaneous  parametric process. The generation modes can in general have distinct spectral properties, which has been emphasized with the well-separated spectra shown on the right.   Finally, another coherent state corresponding to the seed field is included, shown in the figure as $\hat{D}_{seed} (\{ \beta\}) \vac$. The seed field is shown with a wide spectrum to point out that it could overlap with more than one of the TOSPDC generation modes.   Note that we refer to each generation mode which exhibits overlap with the seed field or fields as a seeded mode.  Note also that the light exiting the nonlinear medium at modes $\hat{b}(k_i)$ with $i=\{ 1,2,3\}$ could include spontaneously generated triplets, photons from the applied seed, as well as stimulated radiation resulting from the effect of seeding.   In our analysis, we employ a description of the fields in a dispersive medium \cite{Drummond1999,Drummond1990} and obtain the evolution of the input field asymptotically to the output of the medium in a similar fashion to the treatment in references \cite{Liscidini2012, Liscidini2013}.

In this paper we employ the so-called asymptotic-state formalism \cite{Yang2008,Liscidini2012}, which was developed originally for $\chi^{(2)}$ processes and includes the possibility of multiple pairs of photons per generation event and the use of seeding, to third order parametric downconversion for the first time to the best of our knowledge.    This approach has two important benefits as compared to other published Heisengerg-formalism treatments:  i) the full quantum-mechanical nature of the pump and seed fields can be retained, and ii) it permits the full three dimensional propagation of the fields, including the transverse amplitude, instead of restricting the propagation to a given axis.  Note that while the case studied in this paper is one-dimensional in nature (specific fiber modes), it is useful to have an approach which is ready to be applied in a more general context.

We study the third-order parametric downconversion process with the total Hamiltonian $H= H_L  + H_{NL}$.    $H_L= \hbar \sum_\mu \int dk \omega_\mu(k)  \hat{b}^\dagger (k)  \hat{b}(k)$, with $\mu$ designating each of the  fields involved, is the linear part and the nonlinear part is given as
\cite{Drummond1999a,Drummond1999},

\begin{equation}
	H_{NL} =  \frac{3}{4} \epsilon_0  \int d\ve{r}   \ve{E} \cdot  \boldsymbol\chi^{(3)} \ve{E}\cdot \ve{E} \cdot \ve{E} + h.c.,
	\label{eq:HNL1}
\end{equation}

\noindent where $\epsilon_0$ is the vacuum electric permittivity, $\chi^{(3)}$ is the third-order non-linear susceptibility of the medium and $\ve{E}$ the total electric field, which contains the sum of all fields involved as $\ve{E} = \sum_\mu \ve{E}_\mu$. Individual  fields propagate in the $x$ direction with polarization $\sigma$ and transverse normalized amplitude $u^\perp(y,z)$; they can be written as

\begin{equation}
\begin{aligned}
\ve{E}_\mu (x) &= i \hat{\mathbf{e}}_{\sigma}   \int dk  \sqrt{  \frac{\hbar k_\mu  \nu_{k}  }{4\pi \epsilon(\omega (k_\mu) ) }  }  \\
&\quad \times (u^{\bot}_\mu(y,z)  \hat{b}_\mu (k_\mu) e^{ik_\mu x} - u_\mu^{\bot *}(y,z)\hat{b}^\dagger_{\mu}(k_\mu) e^{-ik_\mu x}  ),
\end{aligned}
\label{eq:Efield}
\end{equation}

\noindent where $\hbar$ is the Planck constant, $k$ the wavenumber, $\nu_k$ the group velocity, $b^\dagger_{\mu}(k) $ is the creation operator for field  $\mu$ and $\epsilon(\omega (k) ) = k^2/\mu_0 \omega^2(k)$ with $\mu_0$ the permeability of free space. In this treatment, we will consider co-polarized and collinear fields, allowing for a scalar description of (\ref{eq:Efield}). The  Schr\"{o}dinger equation for the Hamiltonian is as follows
\begin{equation}
i\hbar \frac{d}{dt} \ket{\psi(t)}= \mathcal{H}_{NL} (t)  \ket{\psi(t)},
\label{eq:SchrodGral}
\end{equation}

\noindent where the time-dependent Hamiltonian is obtained as $ \mathcal{H}_{NL} (t) = \int d^3 \ve{r} e^{H_L t /\hbar} H_{NL} e^{-H_L t /\hbar}$. It can be shown that the energy conserving term for the TOSPDC process is as follows 
\begin{equation}
\begin{aligned}
\mathcal{H}_{NL}(t) &= \int  dk_p \int dk_1 \int dk_2 \int dk_3 S(k_1,k_2,k_3,k_p)\\
&\times \hat{b}^\dagger (k_1) \hat{b}^\dagger (k_2) \hat{b}^\dagger (k_3) \hat{a} (k_p)   e^{-i \Delta \omega t} +h.c.,
\end{aligned}
\vspace{0.01cm}
\label{eq:HNL}
\end{equation}

\noindent where $\Delta \omega = \omega_p-\omega_1-\omega_2-\omega_3$, with $\omega_p$ and  $\omega_i$  (with $i=1,2,3$)  the frequency of the pump and the downconverted fields, respectively, and $h.c.$ denotes the Hermitian conjugate.  In Eq. \ref{eq:HNL} the function $ S(k_1,k_2,k_3,k_p)$ is defined as
\begin{equation}
\begin{aligned}
S&(k_1,k_2,k_3,k_p)= - \frac{9\hbar^2}{32\pi^2} \epsilon_0 \chi^{(3)}\\
&\times \left[\frac{ k_p k_1 k_2 k_3 \nu_{kp} \nu_{k1}\nu_{k2}\nu_{k3}}{ \epsilon(\omega(k_p)) \epsilon(\omega(k_1)) \epsilon(\omega(k_2)) \epsilon(\omega(k_3)) }  \right]^{1/2}\\
&\times \int d^3\ve{r} u^\bot_p(y,z) u^{\bot*}_1(y,z)u^{\bot*}_2(y,z)u^{\bot*}_3(y,z) e^{-i(\Delta k)x},
\end{aligned}
\end{equation}
where $\Delta k = -k_p +k_1+k_2+k_3$, which is also used in the more explicit form  $\Delta k (\omega_1,\omega_2,\omega_3)= -k_p(\omega_1+\omega_2+\omega_3) +k_1(\omega_1)+k_2(\omega_2)+k_3(\omega_3)$ throughout the text. 

We proceed in a similar fashion to reference \cite{Liscidini2013}, with the asymptotic-in fields to the non-linear medium as coherent states in the form
\begin{equation}
\begin{aligned}
\ket{\psi_{in}}&= \exp \left(\int \alpha(k) \hat{a}^\dagger_k dk -h.c. \right) \\ & \times \exp \left(\int \beta(k) \hat{b}^\dagger_k dk -h.c. \right)\vac,
\end{aligned}
\label{eq:psiIn}
\end{equation}
where $\alpha(k), \beta(k)$ are the spectral envelopes of the pump and the seed fields, defined such that $\int dk |\alpha (k) | ^2 (\int dk |\beta (k) | ^2 ) $ represents the average photon number of the pump (seed) in the interaction time. The time dependent solution to (\ref{eq:SchrodGral}) is given as
\begin{equation}
\begin{aligned}
\ket{\psi (t)}&= \exp \left(\int \overline{\alpha}(k,t) \hat{a}^\dagger_k dk -h.c. \right)  \\ &\times \exp \left(\int \overline{\beta}(k,t) \hat{b}^\dagger_k dk -h.c. \right)\ket{\varphi(t)}.
\end{aligned}
\label{eq:psit}
\end{equation}

In the above equation, the resulting complete state of the system  $\ket{\psi (t)}$  is expressed  in such a way that the evolution of the pump and seed fields appears explicitly (in the  operator formed by the product of the two exponentials in Eq. (\ref{eq:psit}); as is discussed below this corresponds to the classical evolution  - i.e. in the absence of quantum mechanical effects - according to a set of coupled differential equations (Eqns. (\ref{eqs:movClassicalFields}), below).   $\ket{\psi (t)}$ is  then be formed by this operator acting on a state $\vert \varphi(t) \rangle$, itself the result of a perturbative calculation (see Eq. (\ref{eq:evol}), below) based on an effective Hamiltonian (see Eq. (\ref{eq:HamiltEff}), below), which will lead to the result in Eq. (\ref{eq:edoModinfty}).    The quantities $\overline{\alpha}(k,t)$ and $\overline{\beta}(k,t)$ in Eq. (\ref{eq:psit}) represent the temporal evolution of the coherent input fields as

\begin{subequations}
\begin{equation}
\overline{\alpha}(k,t) = \alpha (k) + \tilde{\alpha}(k,t),
\end{equation}
\begin{equation}
\overline{\beta}(k,t) = \beta (k) + \tilde{\beta}(k,t).
\end{equation}
\label{eq:alphabetabar}
\end{subequations}

\noindent Note that in Eq. \ref{eq:alphabetabar}, $\tilde{\alpha}(k,t)$ and $\tilde{\beta}(k,t))$ are the time dependent parts of the spectral envelopes. We assume that the pump and seed fields follow a classical evolution description as in reference \cite{Liscidini2013}. The quantum operators for the pump and seed field are substituted by their classical amplitude fields; it is then straightforward to obtain the following coupled set of equations for the amplitudes $\bar{\alpha}$ and $\bar{\beta}$

\begin{subequations}
\begin{equation}
\begin{aligned}
\frac{d \overline{\alpha} (k_p,t)}{dt}& =-\frac{i}{\hbar} \int dk_1 \int dk_2 \int dk_3 S(k_1,k_2, k_3, k_p)\\
&\times  \overline{\beta}(k_1,t) \overline{\beta}(k_2,t) \overline{\beta}(k_3,t) e^{-i\Delta \omega t},
\end{aligned}
\label{eq:movClassicalFieldsAlpha}
\end{equation}
\begin{equation}
\begin{aligned}
\frac{d \overline{\beta} (k_1,t)}{dt}&=-\frac{3i}{\hbar} \int dk_2 \int dk_3 \int dk_p S(k_1,k_2, k_3, k_p)\\
&\times \overline{\beta}^*(k_2,t) \overline{\beta}^*(k_3,t) \overline{\alpha} (k_p,t) e^{-i\Delta \omega t}.
\end{aligned}
\label{eq:movClassicalFieldsBeta}
\end{equation}
\label{eqs:movClassicalFields}
\end{subequations}
Note that in the undepleted pump approximation $d \overline{\alpha} (k_p,t)/dt =0$, therefore $\tilde{\alpha}(k,t)=0$. By substitution of eq. (\ref{eq:psit}) into eq. (\ref{eq:SchrodGral}) one can obtain the effective Hamiltonian
\begin{widetext}
\begin{equation}
\begin{aligned}
H_{eff}(t) &= \int dk_p \int dk_1 \int dk_2 \int dk_3 S(k_1,k_2, k_3, k_p)  \hat{b}^\dagger (k_1) \hat{b}^\dagger (k_2) \hat{b}^\dagger (k_3) \hat{a} (k_p)  e^{-i\Delta \omega t}\\
&\quad + \int dk_p \int dk_1 \int dk_2 \int dk_3 S(k_1,k_2, k_3, k_p)  \hat{b}^\dagger (k_1) \hat{b}^\dagger (k_2) \hat{b}^\dagger (k_3) \alpha (k_p) e^{-i\Delta \omega t}\\
&\quad +3\int dk_p \int dk_1 \int dk_2 \int dk_3 S(k_1,k_2, k_3, k_p)  \overline{\beta}^*(k_1,t) \hat{b}^\dagger (k_2) \hat{b}^\dagger (k_3) \hat{a} (k_p) e^{-i\Delta \omega t}\\
&\quad + 3 \int dk_p \int dk_1 \int dk_2 \int dk_3 S(k_1,k_2, k_3, k_p)  \overline{\beta}^*(k_1,t) \hat{b}^\dagger (k_2) \hat{b}^\dagger (k_3)  \alpha (k_p) e^{-i\Delta \omega t}\\
&\quad + 3 \int dk_p \int dk_1 \int dk_2 \int dk_3 S(k_1,k_2, k_3, k_p)  \overline{\beta}^*(k_1,t) \overline{\beta}^*(k_2,t)  \hat{b}^\dagger (k_3)  \hat{a} (k_p)e^{-i\Delta \omega t} + h.c.,
\end{aligned}
\label{eq:HamiltEff}
\end{equation}
\end{widetext}
which obeys the Schr\"{o}dinger equation $i\hbar \frac{d}{dt} \ket{\varphi(t)} = H_{eff} \ket{\varphi(t)}$.  We can obtain the resulting quantum state through a standard first order time-dependent perturbative approach for  $\ket{\varphi(t)}$ with the aid of eq. (\ref{eq:HamiltEff}); the state at $t\rightarrow\infty$ is obtained as

\begin{eqnarray}\label{eq:evol}
\ket{\varphi(\infty)} &\approx \left(1+\frac{1}{i \hbar}\int\limits_0^\infty d t' H_{eff}(t') \right)\vac
\end{eqnarray}

In what follows, we will assume that the three TOSPDC modes are created in the same spatial mode and with the same polarization, which simplifies the effective Hamiltonian, since the function  $S(k_1,k_2,k_3,k_p)$  becomes symmetric in the sense of being invariant to permutations of the arguments  $k_1$, $k_2$ and $k_3$. 


Note that in  obtaining an expression for the state $\ket{\varphi(\infty)}$ from Eq. \ref{eq:evol}, only 2 out of 5 terms in the effective Hamiltonian, see Eq. \ref{eq:HamiltEff}, (the second and fourth terms) yield a contribution to the resulting quantum state;  the remaining terms vanish because they involve an annihilation operator acting on the vacuum state. The resulting state can be expressed as follows

\begin{equation}
\begin{aligned}
&\ket{\varphi(\infty)} =\vac \\
&\quad + \frac{2\pi}{i\hbar}\int dk_1 \int dk_2 \int dk_3 \int dk_p S(k_1,k_2,k_3,k_p)  \\
&\quad \times \hat{b}^\dagger (k_1) \hat{b}^\dagger (k_2) \hat{b}^\dagger (k_3)  \alpha(k_p)\delta (\omega_p-\omega_1-\omega_2-\omega_3)\vac\\
&\quad+ \frac{6\pi}{i\hbar}\int dk_1 \int dk_2 \int dk_3 \int dk_p S(k_1,k_2,k_3,k_p) \\
&\quad \times \overline{\beta}^* (k_1,t) \hat{b}^\dagger (k_2) \hat{b}^\dagger (k_3)  \alpha (k_p)\delta (\omega_p-\omega_1-\omega_2-\omega_3)\vac,
\end{aligned}
\label{eq:edoModinfty}
\end{equation}

\noindent where the first term is associated with third order spontaneous parametric downconversion (TOSPDC)\cite{Corona2011,Chekhova2005}, and the second term is associated with stimulated, or seeded, third order parametric downconversion (STOPDC).

The state in eq. (\ref{eq:edoModinfty}) can be expressed as
\begin{equation}
\ket{\varphi(\infty)} = \mathscr{N} (\vac + c_{III}\ket{III}  + c_{II} \ket{II} )
\label{eq:phiInf}
\end{equation}

\noindent with $\mathscr{N}=1/\sqrt{1+ c_{III}^2+ c_{II}^2}$, and in terms of a three-photon term  $\ket{III}$ derived from the spontaneous process, as well as a two-photon term  $\ket{II}$ derived from the seeded process, where $ c_{III}$ and $ c_{II}$ are the corresponding probability amplitudes, obtained from Eq. (\ref{eq:edoModinfty}) through normalization of the states $\ket{III}$  and  $\ket{II}$; these states can be written as follows

\begin{subequations}
\begin{equation}
\begin{aligned}
\ket{III}&=\frac{1}{\sqrt{6}} \int dk_1  \int dk_2  \int dk_3 \phi_{III} (k_1,k_2,k_3) \\
&\times\hat{b}^\dagger (k_1) \hat{b}^\dagger (k_2) \hat{b}^\dagger (k_3) \vac,
\end{aligned}
\label{eq:triplete}
\end{equation}
\begin{equation}
\ket{II}=\frac{1}{\sqrt{2}} \int dk_1 \int dk_2  \phi_{II} (k_1,k_2) \hat{b}^\dagger (k_1) \hat{b}^\dagger (k_2) \vac,
\label{eq:biphoton}
\end{equation}
\end{subequations}

\noindent in terms of functions $\phi_{III} (k_1,k_2,k_3)$ and $\phi_{II}  (k_1,k_2)$, which are in turn normalized so that the integral of the absolute value squared over all $k$-number arguments yields unity.  $\phi_{III} (k_1,k_2,k_3)$ and $\phi_{II}  (k_1,k_2)$ can be expressed as follows

\begin{subequations}
\begin{equation}
\begin{aligned}
\phi_{III}(k_1,k_2,k_3) &= \frac{2\sqrt{6}\pi}{i\hbar c_{III}} \int dk_p S(k_1,k_2,k_3,k_p) \alpha(k_p) \\
&\times\delta (\omega_p-\omega_1-\omega_2-\omega_3),
\end{aligned}
\label{eq:phi}
\end{equation}
\begin{equation}
	\phi_{II} (k_1,k_2) = \sqrt{3} \frac{c_{III}}{c_{II}}\int dk_3 \phi_{III} (k_1,k_2,k_3) \beta^*(k_3),
\end{equation}
\label{eq:phiVarphi}
\end{subequations} 
where the terms $c_{II}$ and $c_{III}$ can be expressed as follows

\begin{subequations}
\begin{equation}
\begin{aligned}
|c_{II}|^2 &= \frac{72 \pi^2}{\hbar^2} \int dk_1 dk_2 dk_3 dk_3' dk_p dk_p' S^*(k_1,k_2,k_3',k_p')\\
&\quad\times S(k_1,k_2,k_3,k_p) \beta(k_3') \beta^*(k_3) \alpha^*(k_p') \alpha(k_p) \\
&\quad\times\delta (\omega_p'-\omega_1-\omega_2-\omega_3')\delta (\omega_p-\omega_1-\omega_2-\omega_3),
\end{aligned}
\end{equation}
\begin{equation}
\begin{aligned}
|c_{III}|^2 &= \frac{24 \pi^2}{\hbar^2} \int dk_1 dk_2 dk_3 dk_p dk_p' S^*(k_1,k_2,k_3,k_p)\\
&\quad\times S(k_1,k_2,k_3,k_p') \alpha^*(k_p) \alpha(k_p')\\
&\quad\times \delta (\omega_p-\omega_1-\omega_2-\omega_3)\delta (\omega_p'-\omega_1-\omega_2-\omega_3).
\end{aligned}
\end{equation}
\end{subequations}

In the undepleted pump approximation for which both sides of eq.  (\ref{eq:movClassicalFieldsAlpha}) vanish, we can integrate Eq.  (\ref{eq:movClassicalFieldsBeta}) so as to obtain 
the following expression for $\overline{\beta}(k_1,t)$, in terms of function  $\phi_{III}(k_1,k_2,k_3)$

\begin{eqnarray}
&\overline{\beta}(k_1,t) = \beta(k_1)+  \nonumber \\
&\sqrt{\frac{3}{2}} c_{III} \int dk_2 \int dk_3 \phi_{III}(k_1,k_2,k_3) \beta^*(k_2) \beta^*(k_3).
\label{eq:betabarra}
\end{eqnarray}

Within this approximation,  Eq.(\ref{eq:psit}) can then be written in terms of (\ref{eq:betabarra}) as 

\begin{equation}
\begin{aligned}
&\ket{\psi_{out}}  = \hat{D}(\{ \alpha_p \})  \exp \left( \int\left(   \overline{\beta}(k_1,t) \right)\hat{b}^\dagger (k_1)dk_1  -h.c.       \right)  \\
&\quad\times \mathscr{N} (\vac+c_{III} \ket{III} + c_{II} \ket{II}),
\end{aligned}
\label{eq:psiOut}
\end{equation}

\noindent where $\hat{D}(\{ \alpha_p \}) $ is the displacement operator for the pump.  Note that we assume that the spectral overlap between the pump and the TOSPDC generation modes is negligible, a reasonable assumption  considering the large spectral separation between them.

It is interesting to point out that equivalent expressions for the stimulated process can be obtained if one starts from a description of the electric field for each TOSPDC generation mode as follows

\begin{equation}
\tilde{\ve{E}}_i(x,t)= \ve{\hat{E}}_i(x,t) + \ve{E}_i^{\text{cl}} (x,t),
\label{eq:EQuantClassField}
\end{equation}

\noindent where $\ve{\hat{E}}_i(x,t) $ is the corresponding quantized electric field, and $\ve{E}_i^{\text{cl}} (x,t)$ is the classical seed field. By substitution of  Eq. (\ref{eq:EQuantClassField}) into  the nonlinear Hamiltonian in Eq. (\ref{eq:HNL1}), we obtain three energy conserving terms as follows

\begin{equation}
\begin{aligned}
H_{NL}(t)&= H_{\text{0}}(t) + H_{1} (t) + H_{2} (t),
\end{aligned}
\label{eq:H_TOPDC}
\end{equation}

\noindent where $H_{\text{0}}(t)$ represents the spontaneous process, while $H_{1} (t)$ and $H_{2} (t)$ include the effect of the seed field overlapping with one or two TOSPDC modes, respectively.   These terms can be expressed as follows 

\begin{subequations}
	\begin{equation}
	\begin{aligned}
	H_{\text{0}}(t) &= \int dk_p \int  dk_1  \int dk_2  \int dk_3 S(k_1,k_2,k_3,k_p)\\
	&\times \hat{b}^\dagger (k_1) \hat{b}^\dagger (k_2) \hat{b}^\dagger (k_3) \hat{a} (k_p)   e^{-i \Delta \omega t} +h.c.,
	\end{aligned}
	\vspace{0.01cm}
	\label{eq:H_TOSPDC}
	\end{equation}
	\begin{equation}
	\begin{aligned}
	H_{1}(t)&=3 \int dk_p  \int dk_1  \int dk_2  \int dk_3 S(k_1,k_2,k_3,k_p)  \\
	&\times \beta^* (k_3) \alpha(k_p) \hat{b}^\dagger(k_1) \hat{b}^\dagger(k_2)  e^{-i \Delta \omega t} +h.c.,
	\end{aligned}
	\label{eq:H_I} 
	\end{equation}
	\begin{equation}
	\begin{aligned}
	H_{2}(t)&=3 \int dk_p  \int dk_1  \int dk_2  \int dk_3 S(k_1,k_2,k_3,k_p)  \\
	&\times \beta^* (k_2) \beta^* (k_3) \alpha(k_p) \hat{b}^\dagger(k_1)   e^{-i \Delta \omega t} +h.c.
	\end{aligned}
	\label{eq:H_II}
	\end{equation}
\end{subequations}

Through the standard perturbative approach to first order, we obtain the following state 

\begin{equation}
\ket{\Psi_{out}} = \mathscr{N}' (\vac + c_{III}  \ket{III} + c_{II} \ket{II} + c_{I} \ket{I}),
\label{eq:PsiOut}
\end{equation}

\noindent in terms of normalization constant $\mathscr{N}'$, where the expressions for $\ket{III}$ and $\ket{II}$ are identical to those found above, see Eqns. (\ref{eq:triplete}) and (\ref{eq:biphoton}), while $\ket{I}$ is expressed as follows

\begin{equation}
\ket{I}= \int dk_1  \phi_I (k_1) \hat{b}^\dagger (k_1) \vac,
\end{equation}

\noindent with

\begin{equation}
\begin{aligned}
\phi_I(k_1) &= \sqrt{\frac{3}{2}} \frac{c_{III}}{c_I} \int dk_2 dk_3 \phi(k_1,k_2,k_3) \beta^*(k_2)\beta^*(k_3).
\end{aligned}
\label{eq:Theta}
\end{equation}

Note the similarity of Eq. (\ref{eq:Theta}) with the second term in Eq. (\ref{eq:betabarra}).  Also note that through this approach we obtain directly a one-photon contribution derived from dual seeding, while in the previous asymptotic treatment this contribution appears implicitly. In order to obtain the double seeded contribution in the asymptotic treatment, one may expand the exponential terms in (\ref{eq:psit}), where it should be noted that  the seed operators do not commute with those associated with the TOSPDC generation modes.  Substitution of Eq. (\ref{eq:phiInf}) into Eq. (\ref{eq:psit}) then yields the following output state

\begin{equation}
\begin{aligned}
\ket{\psi(\infty)} &=  \exp \left(\int \overline{\alpha}(k,t) \hat{a}^\dagger_k dk -h.c. \right) \\
&\times \left[ 1+ \int dk \overline{\beta}(k,t)\hat{b}^\dagger (k) -h.c. +\cdots \right]\\
&\times \left[ \vac + c_{III}\ket{III} + c_{II} \ket{II}  \right], 
\end{aligned}
\end{equation}

\noindent and by using the expression for $\overline{\beta}(k,t)$ obtained in Eq. (\ref{eq:betabarra}) one  obtains the one photon (double seeded)  term with the correct coefficient.  Of course, higher-order terms which we will ignore here appear when expanding the exponential  in eq. (\ref{eq:psit}).

\section{Emitted photon flux}

We are interested in calculating the effect of seeding on the observed intensities of the output modes, as a guide for future experiments.   To this end, we calculate the expectation value of the number operator for one of the output modes, integrated over all $k$ wavenumbers as follows

\begin{equation}
N= \int dk \bra{\psi_{out}}   \hat{b}^\dagger (k) \hat{b}(k) \ket{\psi_{out}},
\label{eq:NPhotons}
\end{equation}
where $N$ is calculated within the interaction time, which is taken as the pulse duration if at least one of the fields (pump and seed) is pulsed, or as the unit time (1s)  if all fields are CW. To obtain expectation values of the number operator per unit time in pulsed cases, the expression in (\ref{eq:NPhotons}) should be multiplied by the repetition rate $R$.

\noindent  By substitution of (\ref{eq:psiOut}) into (\ref{eq:NPhotons}) and assuming that the generated modes commute with the pump modes, we can group the resulting expression in terms of the number of modes that  overlap the seed as follows

\begin{equation}
N=N_{0}+N_{1}+N_{2},
\label{eq:NPhotons2}
\end{equation} 

\noindent where the Baker-Campbell-Hausdorff \cite{Scully10} formula has been used so as to rearrange the non-commuting terms related to the observed mode and the seed modes as $\hat{D}^\dagger (\{ \overline{\beta}(k_1,t) \}) \hat{b}(k) \hat{D} (\{ \overline{\beta}(k_1,t) \}) =\hat{b}(k_1)+\beta(k_1)+ \sqrt{\frac{3}{2}}c_{III} \int dk_2dk_3 \phi(k,k_2,k_3) \beta^*(k_2) \beta^*(k_3) $.  Note that in all calculations the integral interval will not include the seed spectral range, to avoid summing the expectation value of the seed  photon number to the stimulated photons. The term $N_{0}$ in eq. (\ref{eq:NPhotons2}) corresponds to the rate of spontaneous triplet generation, while the terms $N_1$ and $N_{2}$ correspond to the generation rate in the TOSDPC modes,  given the presence of a seed field that can overlap one or two of the existing output modes, respectively. The spontaneous term is easily evaluated as 

\begin{equation}
N_{0} = 3 |c_{III}|^2,
\label{eq:Ntospdc}
\end{equation}

\noindent while the terms $N_{1}$ and $N_{2}$ are obtained from eq. (\ref{eq:NPhotons}) with the aid of eq. (\ref{eq:psiOut}) as follows

\begin{subequations}
\begin{equation}
\begin{aligned}
N_1&=2 N_0 |\beta_0|^2  \Theta_1,
\end{aligned}
\end{equation}
\begin{equation}
\begin{aligned}
N_{2}&=\frac{N_0}{2} |\beta_0|^4   \Theta_2,
\end{aligned}
\end{equation}
\label{eq:NI_NII}
\end{subequations}

\noindent  where we have written $\beta(k)$ as  $\beta(k)=\beta_0 \tilde{\beta}(k)$;  here, $\tilde{\beta}(k)$ is normalized so that $|\tilde{\beta}(k)|^2 $ has a unit integral, while $|\beta_0|^2 $ represents the average photon number of the seed field, in terms of the frequency-integrated single-seed and dual-seed overlap terms  $\Theta_1$ and $\Theta_2$

\begin{subequations}\label{Eq:overlap}
	\begin{equation}
\Theta_1= \int d k_1  \int dk_2 \left|\int dk_3 \phi_{III}  (k_1,k_2,k_3) \tilde{\beta}^*(k_3) \right|^2,  
	\end{equation}
\begin{equation}
\Theta_2=  \int d k_1\  \left|  \int dk_2  \int  dk_3 \phi_{III} (k_1,k_2, k_3) \tilde{\beta}^*(k_2) \tilde{\beta}^*(k_3)    \right|^2.
\end{equation}
\end{subequations}

It becomes clear that  the resulting flux is proportional to the product of three terms: i) the spontaneous, unseeded flux, ii) the seed intensity (square of the seed intensity) for the single-seed (double-seed) case, and iii) a frequency-integrated spectral overlap term between the three-photon amplitude function $ \phi_{III}  (k_1,k_2,k_3)$ and the seed.  $N_1$  quantifies the flux produced by the singly-seeded process which is mathematically described by the second term of Eq. \ref{eq:edoModinfty}.  Note that this contribution can be understood as analogous to the quantum state produced by the processes of spontaneous parametric dowconversion (SPDC) or spontaneous four wave mixing (SFWM), in which for a sufficiently low parametric gain photon pairs are produced; such a process  has no classical analogue.    In contrast, the flux represented by $N_2$ derived from the presence of two seeds can be fully understood in terms of the classical equations of motion for the pump and  seed fields, see Eqns. (\ref{eqs:movClassicalFields}a) and (\ref{eqs:movClassicalFields}b), and corresponds to classical difference frequency generation in which a new field with frequency $\omega_3 = \omega_p-\omega_1-\omega_2$ is generated, with $\omega_p$ the pump frequency and $\omega_1$ and $ \omega_2$ the seed frequencies.  Note also that the singly-seeded case can be understood as double seeding with one of the seeds corresponding to vacuum fluctuations.


It is clear from equations (\ref{Eq:overlap})  that the effect of seeding will be highly dependent on the spectrum of the seed. Note that a significant difference arises compared to the photon-pair case studied by Liscidini and Sipe \cite{Liscidini2013}, where only  the term equivalent to our $N_1$ exists. 

While we have shown  expressions for the total flux (integrated over all wavenumbers of the mode in question), on occasion it is the emission spectra instead which are of interest.   We thus define the singly-overlapped and doubly-overlapped emission spectra $N_1(k_1)$ and $N_2(k_1)$, respectively,  so that $\int d k_1 N_1(k_1)= N_1$ and $\int d k_1 N_2(k_1)= N_2$.  This leads to the following expressions.

\begin{align}
N_1(k_1)&=2 N_0 |\beta_0|^2 \Theta_1(k_1),   \\ \nonumber 
N_{2}(k_1)&=\frac{N_0}{2} |\beta_0|^4  \Theta_2(k_1),
\end{align}

\noindent in terms of spectrally-resolved overlap coefficients $\Theta_1(k_1)$ and  $\Theta_2(k_1)$, which are defined from eq. (\ref{Eq:overlap})

\begin{align}
\Theta_1(k_1)&= \int dk_2 \left|\int dk_3 \phi_{III}  (k_1,k_2,k_3) \tilde{\beta}^*(k_3) \right|^2,
\end{align}

\begin{align}
\Theta_{2}(k_1)&=  \left|  \int dk_2  \int  dk_3 \phi_{III} (k_1,k_2, k_3) \tilde{\beta}^*(k_2) \tilde{\beta}^*(k_3)    \right|^2.
\end{align}

\subsection{Case I: Spontaneous, unseeded process}

We proceed to calculate the photon flux in the absence of a seed, i.e. for $\beta(k)=0$, so as to establish a link with previous TOSPDC studies \cite{Corona2011,Corona2011a} and so as to define the expressions that will be used for the seeded cases.

It may be shown that the coefficient $|c_{III}|^2$, which determines the spontaneous generation rate as $N_0=3 |c_{III}|^2$, may be expressed as follows in the case of a pulsed pump

\begin{equation}
\begin{aligned}
&|c_{III}|^2_{pulsed}= \frac{3^3\sqrt{2} \hbar L^2  n_0^4 P_{av}}{8\pi^{5/2} \omega_0^2\sigma_p R} |\gamma|^2\\
&\times  \int d\omega_1 \int d\omega_2 \int d\omega_3 \frac{\omega_1 \omega_2 \omega_3 |f(\omega_1,\omega_2,\omega_3)|^2}{n(\omega_1) n(\omega_2) n(\omega_3) n(\omega_1 + \omega_2 + \omega_3)},
\end{aligned}
\label{eq:gamma}
\end{equation}

\noindent in terms of the the repetition rate $R$,  average power $P_{av}$, and bandwidth $\sigma_p$ of the pump laser;  $L$ the length of the non-linear medium, $n(\omega_i)$  is the refractive index at frequency $\omega_i$   (with $i=1,2,3$), $n_0$ the refractive index at the central pump frequency $\omega_0$.   The non-linear coefficient $\gamma$ can be expressed as \cite{Garay-Palmett2010}

\begin{equation}
\gamma = \frac{3\chi^{(3)} \omega_o f_{eff} }{4\epsilon_0 c^2 n_0^2},
\end{equation}

\noindent written in terms of the spatial overlap $f_{eff}$ between the four fields involved in the TOSPDC and STOPDC processes

\begin{equation}
f_{eff} = \int_{-\infty}^{\infty} dy \int_{-\infty}^{\infty} dz u^\bot_p(y,z) u^{\bot*}_1(y,z)u^{\bot*}_2(y,z)u^{\bot*}_3(y,z).
\end{equation}

In Eq. \ref{eq:gamma}, the joint amplitude function $f(\omega_1,\omega_2,\omega_3) $ can be written as
\begin{equation}
\begin{aligned}
f(\omega_1,\omega_2,\omega_3) &= \xi(\omega_1,\omega_2,\omega_3)  \Xi (\omega_1,\omega_2,\omega_3),\\
\end{aligned}
\end{equation}

\noindent with

\begin{equation}
\begin{aligned}
\xi(\omega_1,\omega_2,\omega_3)& =  e^{-(\omega_1+\omega_2+\omega_3-\omega_0)^2/\sigma_p^2}, \\ 
\Xi (\omega_1,\omega_2,\omega_3)&= \text{sinc}\left(\frac{L}{2}\Delta k \left(\omega_1,\omega_2,\omega_3\right)\right),
\end{aligned}
\end{equation}

\noindent where the functions $\xi(.)$ and $\Xi(.)$ are the pump envelope and the phase-matching function, respectively.  It is likewise of interest to evaluate the limit  $\sigma_p \to 0$ in Eq. \ref{eq:gamma} so as to obtain the spontaneous generation rate for a monochromatic pump.  It may be shown that in this limit we obtain the following expression for $|c_{III}|^2$

\begin{equation}
\begin{aligned}
&|c_{III}|^2_{cw}=  \frac{3^3 \hbar L^2  n_0^4 P_{av}}{8\pi^2 \omega_0^2} |\gamma|^2 \int d\omega_1 \int d\omega_2\\
&\times  \frac{\omega_1 \omega_2 (\omega_0-\omega_1-\omega_2) |f(\omega_1,\omega_2, \omega_0-\omega_1-\omega_2)|^2}{n(\omega_1) n(\omega_2) n( \omega_0-\omega_1-\omega_2) n(\omega_0)}.
\end{aligned}
\label{eq:Ntospdc_cw}
\end{equation}

The spontaneous contribution to the photon flux $N_0$  can be  obtained with eq. (\ref{eq:gamma}) and eq. (\ref{eq:Ntospdc_cw}) by simple substitution into eq. (\ref{eq:Ntospdc}). The spontaneous case is not explored further in this work, since it has been studied in Refs. \cite{Corona2011,Corona2011a}.   Also, note that in a seeded scenario the spontaneous contribution to the overall flux  will tend to be negligible when compared to the seeded output fields; expressions for the generation rate in the presence of a seed field will be presented in  the following subsections. We divide our analysis according to the spectral properties of the seed. 

\subsection{Case II:  Pulsed seed}

In this subsection, we analyze the case of a pulsed seed, while the pump field is allowed to be pulsed or monochromatic. 

\subsubsection{Case IIa: Pulsed seed and pulsed pump}

We first  analyze the case  for which both pump and seed are pulsed.   Note that seeding will produce an appreciable effect only in those situations for which the pump and seed are temporally and spectrally overlapped.      In the case where both seed and pump are in the form of a train of pulses, this translates into the need for the two trains to: i) be characterized by the same repetition rate, and ii) for the pump and seed pulse maxima to be temporally-coincident, i.e. with a vanishing temporal delay;  in what follows, $t_0$ denotes the temporal delay between the two pulse trains.
It should be pointed out that in practice it may be challenging for the pump and seed pulse trains, at very different frequencies, to be temporally matched.

The spectral envelope of the seed field, assumed to be Gaussian, may then be expressed in terms of the seed central frequency $\omega_{s0}$ and bandwidth $\sigma_s$ as

\begin{equation}
\beta(\omega,t_0) = \beta_0 \left(\frac{2}{\sigma_s^2 \pi} \right)^{1/4} \beta'(\omega) e^{i\omega t_0} 
\end{equation}

\noindent where $\beta'(\omega)=e^ {-(\omega-\omega_{s0})^2/\sigma_s^2}$ represents the adimensional Gaussian  spectral envelope function for the seed. 

The resulting spectral overlap terms $\Theta_1^{p,p}$ and $\Theta_2^{p,p}$, for both fields pulsed (pump and seed)  are, then, as follows 

\begin{subequations}\label{eq:nSeed1_2}
\begin{equation}
\begin{aligned}
\Theta_1^{p,p} &=  \frac{3^3 n_0^3P_{av}L^2 \hbar }{2^2 \pi^3\sigma_p\sigma_s\omega_0^2R_p }\frac{|\gamma|^2}{|c_{III}|^2} \int d \omega_1\\
&\times  \int  d\omega_3 \frac{\omega_1\omega_3}{n(\omega_1)n(\omega_3)} \bigg| \int d\omega_2 \sqrt{\frac{\omega_2}{n(\omega_2)}} \beta^{'*} (\omega_2) e^{-i\omega_2t_0}\\
&\times f(\omega_1,\omega_2,\omega_3) e^{i\frac{L}{2}\Delta k(\omega_1,\omega_2,\omega_3) }\bigg|^2,
\end{aligned}
\end{equation}

\noindent and 

\begin{equation}
	\begin{aligned}
	\Theta_{2}^{p,p}&=
	 \frac{3^3 n_0^3 P_{av}  L^2  \hbar }{2\sqrt{2}  \omega_0^2 \sigma_s^2 \sigma_p \pi^{7/2} R_p} \frac{|\gamma|^2}{|c_{III}|^2} \int d\omega_1 \frac{\omega_1}{n(\omega_1) }\\
	& \times  \bigg| \int d\omega_2 \int  d\omega_3 e^{i \frac{L }{2} \Delta k(\omega_1,\omega_2,\omega_3)}\\
	& \times \sqrt{\frac{\omega_2\omega_3}{n(\omega_2) n(\omega_3)} }  \beta_s^{'*}(\omega_2) \beta_s^{'*}(\omega_3)\\
	& \times f(\omega_1,\omega_2,\omega_3) e^{-i(\omega_2+\omega_3)t_0}  \bigg|^2,
	\end{aligned}\end{equation}
\end{subequations}

\noindent for the cases where the seed overlaps one TOSPDC mode, and two TOSPDC modes, respectively.

\subsubsection{Case IIb: Pulsed seed and monochromatic pump}

From the expressions which appear in the last subsection, it is possible to obtain versions for a monochromatic pump by taking the limit $\sigma_p \to 0$.  We thus obtain the following expressions for the overlap terms $\Theta_1^{cw,p}$ and $\Theta_2^{cw,p}$ valid for a pulsed seed and monochromatic pump

\begin{subequations}
	\begin{equation}
	\begin{aligned}
	&\Theta_1^{cw,p}=\frac{3^3 |\alpha_p|^2 \hbar^2  n_0^3 L^2}{2^2\sqrt{2}\sigma_s\pi^{5/2}\omega_0}\frac{|\gamma|^2}{|c_{III}|^2}\\
	& \times \int d\omega_1 d\omega_3 \frac{\omega_1(\omega_0-\omega_1-\omega_3)\omega_3}{n(\omega_1)n(\omega_0-\omega_1-\omega_3)n(\omega_3)}\\
	&\times |\beta^{'*}(\omega_0-\omega_1-\omega_3)|^2 \Xi^2 (\omega_1, \omega_0-\omega_1-\omega_3,\omega_3) 
	\end{aligned}
	\end{equation}

\noindent and

	\begin{equation}
	\begin{aligned}
	&\Theta_{2}^{cw, p}=2\left(\frac{3}{2}\right)^3\frac{|\alpha_p|^2 \hbar^2  n_0^3 L^2}{\sigma_s^2\pi^{3}\omega_0}\frac{|\gamma|^2}{|c_{III}|^2}\\
	&\times \int d\omega_1 \Bigg|  \int d\omega_3 \left[\frac{\omega_1(\omega^0-\omega_1-\omega_3)\omega_3}{n(\omega_1)n(\omega_0-\omega_1-\omega_3)n(\omega_3)}\right]^{1/2} \\
	&\times\beta^{'*}(\omega_0-\omega_1-\omega_3)  \beta^{'*}(\omega_3)  \Xi (\omega_1, \omega_0-\omega_1-\omega_3,\omega_3)  \Bigg|^2,
	\end{aligned}
	\end{equation}
	\label{eq:CWPulsed}
\end{subequations}

\noindent for the cases where the seed overlaps one TOSPDC mode, and two TOSPDC modes, respectively, where $|\alpha_p|^2$ represents the average pump photon number which temporally overlaps the seed pulse.

\subsection{Case III: Monochromatic seed}

In this section we present expressions for the case where the seed is monochromatic, while the pump is allowed to be either pulsed or monochromatic.

\subsubsection{Case III a: Monochromatic seed and pulsed pump}

 We obtain the following expressions for the overlap terms $\Theta_1^{p,cw}$ and $\Theta_2^{p,cw}$ valid for a pulsed pump and monochromatic seed.

\begin{subequations}
\begin{equation}
\begin{aligned}
\Theta_{1}^{p,cw}&=\frac{3^3\hbar P_{av} n_0^3 L^2 \omega_1 \omega_s'}{2^2\sqrt{2}\pi^{5/2} \sigma_p R\omega_0^2 n(\omega_1)n(\omega_s')} \frac{|\gamma|^2}{|c_{III}|^2}\\
&\times\int \int  d\omega_1 d\omega_2 \frac{\omega_2}{n(\omega_2)} |f(\omega_1,\omega_2,\omega_s')|^2,
\end{aligned}
\end{equation}

and

\begin{equation}
\begin{aligned}
&\Theta_{2}^{p,cw}=\sqrt{2} \left( \frac{3}{2} \right)^3 \frac{P_{av}}{\pi^{5/2} \sigma_p R \omega_0^2} \hbar n_0^3  L^2 \frac{|\gamma|^2}{|c_{III}|^2}\\
&\quad\times \int d \omega_1 |f(\omega_1,\omega_s', \omega_s')|^2 \frac{\omega_1 \omega_s'^2}{n(\omega_1) n^2(\omega_s')},
\end{aligned}
\end{equation}
\end{subequations}

\noindent for the cases where the seed overlaps one TOSPDC mode, and two TOSPDC modes, respectively, where $\omega_s'$ is the frequency of the monochromatic seed.

\subsubsection{Case III b: Monochromatic seed and monochromatic pump}

 We obtain the following expressions for the overlap terms $\Theta_1^{cw,cw}$ and $\Theta_2^{cw,cw}$ valid for the case where both pump and seed are monochromatic.

\begin{subequations}
\begin{equation}
\begin{aligned}
&\Theta_{1}^{cw,cw}=\frac{3^3 \hbar P_{av}  n_0^3 L^2 }{2^3 \pi^{2} \omega_0^2 } \frac{|\gamma|^2}{|c_{III}|^2}\\
&\quad \times \int d\omega_1 \frac{\omega_1 \omega_s' (\omega_0-\omega_1-\omega_s')}{n(\omega_1)n(\omega_s') n(\omega_0-\omega_1-\omega_s')} \\
&\quad \times \Xi^2 (\omega_1, \omega_0-\omega_1-\omega_s',\omega_s'),
\end{aligned}
\end{equation}
\begin{equation}
\begin{aligned}
&\Theta_{2}^{cw,cw}=\left(\frac{3}{2}\right)^3 \frac{\hbar P_{av} n_0^3 L^2  (\omega_0-2\omega_s') {\omega_s'}^2}{\pi^{2} \omega_0^2 n(\omega_0-2\omega_s')n^2(\omega_s')} \frac{|\gamma|^2}{|c_{III}|^2}\\
&\quad \times \Xi^2 (\omega_1, \omega_0-\omega_1-\omega_s',\omega_s').
\end{aligned}
\end{equation}
\label{eq:NIspCWCW}
\end{subequations}

This case for which the pump and seed are both monochromatic, and therefore continuous wave, is particularly interesting because in contrast with the case where both fields are pulsed, temporal overlap between them is guaranteed with no additional effort.   

\subsection{Case IV: Multiple seed fields}

We can straightforwardly extend our analysis to the case where multiple seed fields are simultaneously present, by an appropriate rewriting of the single-seed description.  Let us assume that each seed field is a coherent state described by its respective displacement operator  $\hat{D} (\{\beta_i \})\vac$.  If $n$ different coherent states are superimposed, for example by means of $n-1$ dichroic mirrors, the resulting field can be described as the product $\Pi_i \hat{D} (\{\beta_i \}) \vac $.  This can be simplified under the assumption that there is no spectral overlap between any  two seed fields, i.e.  $\int dk \beta_i^*(k) \beta_j(k) = 0$, for all $i\neq j$, leading to an effective single seed with amplitude
\begin{equation}
\beta(k)\rightarrow  \sum_i \beta_i(k).
\label{eq:MultipleSeeds}
\end{equation}

Note that under these assumptions,  $\int dk |\beta(k)|^2 = \sum_i |\beta_{i}|^2$, where $|\beta_{i}|^2$ is the average photon number for each seed field.  Note also that in the symmetric case for which the  function $\phi(k_1,k_2,k_3)$ is invariant under permutations of its arguments, the seeded throughputs can be expressed as 

\begin{subequations}
\begin{equation}
\begin{aligned}
N_1&\approx 2 N_0 \sum_i \int d k_1  \int dk_2 \left|\int dk_3 \phi (k_1,k_2,k_3) \beta_i^*(k_3) \right|^2, \\
\end{aligned}
\end{equation}
\begin{equation}
\begin{aligned}
N_{2}&\approx N_0 \sum_{i\neq j} \int d k_1 \\
&\quad\times \left|  \int dk_2  \int dk_3 \phi(k_1,k_2, k_3) \beta_i^*(k_2) \beta_j^*(k_3)    \right|^2 \\
& +  \frac{N_0}{2}\sum_{i} \int d k_1 \\
&\quad\times \left|  \int dk_2  \int dk_3 \phi(k_1,k_2, k_3) \beta_i^*(k_2) \beta_i^*(k_3)    \right|^2.
\end{aligned}
\end{equation}
\label{eq:NI_NII_multipleSeeds}
\end{subequations}

One may note that the first term in $N_{2}$, which corresponds to non-degenerate seed fields, can be used as the basis for  tomographic reconstruction if seeds $i,j$ are scanned within the phase-matched interval (see next section).   Note that in the specific situation where there are two seed fields  present, there are in general six contributions to the output field that can be described as follows:

\begin{enumerate}[\itshape i]
	\item spontaneous term, 
	\item single overlap, of seed 1 with one TOSPDC mode. 
	\item double overlap, of seed 1 with two TOSPDC modes. 
	\item single overlap, of seed 2 with one TOSDPC mode. 
	\item double overlap, of seed 2 with two TOSDPC modes.	
	\item double overlap, of seeds 1 and 2, each with a distinct TOSPDC mode.
\end{enumerate}

The possible dominance of some terms over the others will depend on specific configurations of seeds, pump and the nonlinear characteristics of the non-linear medium. 

\subsection{Stimulated Emission Tomography}

Let us consider a specific multiple-seed configuration (see previous section), specifically with two distinct seed fields.  Let us further assume that these seed fields are sufficiently narrow, in frequency (or $k$-number), so that we may approximate the integrals in eqs. (\ref{Eq:overlap}) as

\begin{subequations}
\begin{equation}
\begin{aligned}
\Theta_1 &\approx \int dk_1 \int dk_2 |\phi (k_1,k_2,k_0^i)|^2 \delta k_i,
\end{aligned}
\label{eq:gralCWSimp}
\end{equation}
\begin{equation}
\begin{aligned}
\Theta_2  &\approx \int dk_1 |\phi (k_1,k_0^i,k_0^j)|^2 \delta k_i \delta k_j,
\end{aligned}
\label{eq:gralCWSimp2}
\end{equation}
\end{subequations}

\noindent where $\delta k_i$ ($\delta k_j$) is the $k$-number bandwidth for the seed field $i$ ($j$).  Let us rewrite the expressions for the single-seed and double-seed throughput, Eqs. (\ref{eq:NI_NII}),  explicitly in terms of the two distinct seed fields $i$ and $j$

\begin{subequations}
\begin{equation}
N_1 =2  N_0  |\beta_{k_0^i}|^2 \Theta_1,
\end{equation}
\begin{equation}
N_2 = \frac{N_0}{2}  |\beta_{k_0^i}|^2|\beta_{k_0^j}|^2 \Theta_2,
\end{equation}
\label{eq:N1N2STOPDC}
\end{subequations}
\noindent where $|\beta_{k_0^i}|^2$ is the average photon number of the $i$ seed field, centered at $k_0^i$.

Now, for the double-seed term $N_2$ let us assume that a spectrally-resolved measurement is carried out so as to determine the throughput for seed fields $i$ and $j$ at each wavenumber $k$, $N_2^{ij}(k)$, defined so that $\int d k N_2^{ij}(k)= N_2$.  We likewise define a spectrally-resolved double-seed overlap $\Theta_2(k)$ so that $\int d k \Theta_2(k) = \Theta_2$.  Note that the spectrally-resolved, doubly-seeed overlap term essentially corresponds to the joint spectral amplitude of the TOSPDC triplets as follows $\Theta_2(k_1)=|\phi (k_1,k_0^i,k_0^j)|^2 \delta k_i \delta k_j$.   We can then write, 

\begin{align}
N_2^{i j}(k_1) &= \frac{N_0}{2}  |\beta_{k_0^i}|^2|\beta_{k_0^j}|^2 \Theta_2(k_1)\nonumber \\
&=\frac{N_0}{2}  |\beta_{k_0^i}|^2|\beta_{k_0^j}|^2 |\phi (k_1,k_0^i,k_0^j)|^2 \delta k_i \delta k_j.
\label{eq:N1N2comp}
\end{align}

It is then a simple matter to re-write this expression as

\begin{equation}
\frac{N_0}{2} |\phi (k_1,k_0^i,k_0^j)|^2  = \frac{N_2^{ij}(k_1)}{ |\beta_{k_0^i}|^2|\beta_{k_0^j}|^2 \delta k_i \delta k_j}.
\end{equation}

This relationship  forms the basis for the stimulated emission tomography (SET) which could be implemented for the spectral characterization of the photon-triplet joint spectral intensity.   For each pair of values $k_i$ and  $k_j$ which are scanned (rasterized)  within the phasematched region, the resulting measured spectrum $N_2^{ij}(k_1)$ is divided by the product of the seed intensities and their bandwidths.   By accumulating measurements for the different $k_i$ and $k_j$ values, one may in principle extract the desired $ |\phi (k_1,k_0^i,k_0^j)|^2$ joint spectral intensity for the TOSPDC photon triplets.

Note that while photon-triplet SET is based on two independent, singly-overlapped seed fields, the presence of additional signals derived from i) doubly-overlapped seeding for one or both of the seed fields, and ii) one of the seed fields exhibiting overlap but not the other, may constitute sources of noise for the SET measurement, since it is impossible to discern whether a particular output photon is derived from SET or from these two other competing seeded TOSPDC variations.   However, SET is likely to yield usable information because: i)the signal obtained from a doubly-overlapped single seed tends to be spectrally localized, and ii) the signal from single seeding is orders of magnitude smaller than that derived from double seeding. 


\section{Stimulated generation in a specific situation}

\begin{figure}
\includegraphics[width=0.35\textwidth]{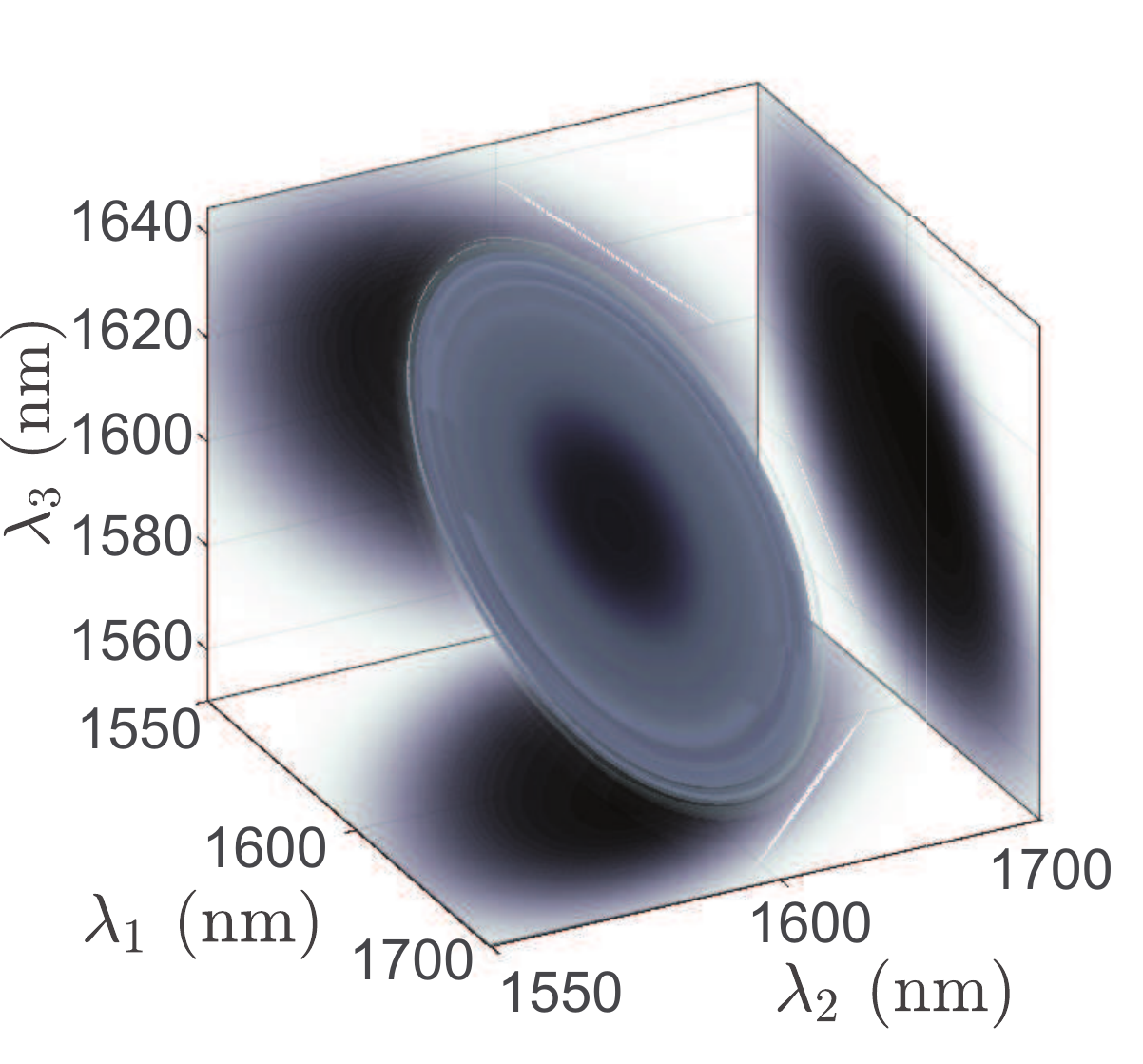}
\caption{Plot of the TOSPDC joint spectral intensity $|f(\lambda_1,\lambda_2,\lambda_3)|^2$ in the frequency-degenerate source configuration.   Marginal distributions are shown on each of the three coordinate planes.}
\label{Fig:JSITripletes}
\end{figure}

\begin{figure}
\includegraphics[width=0.35\textwidth]{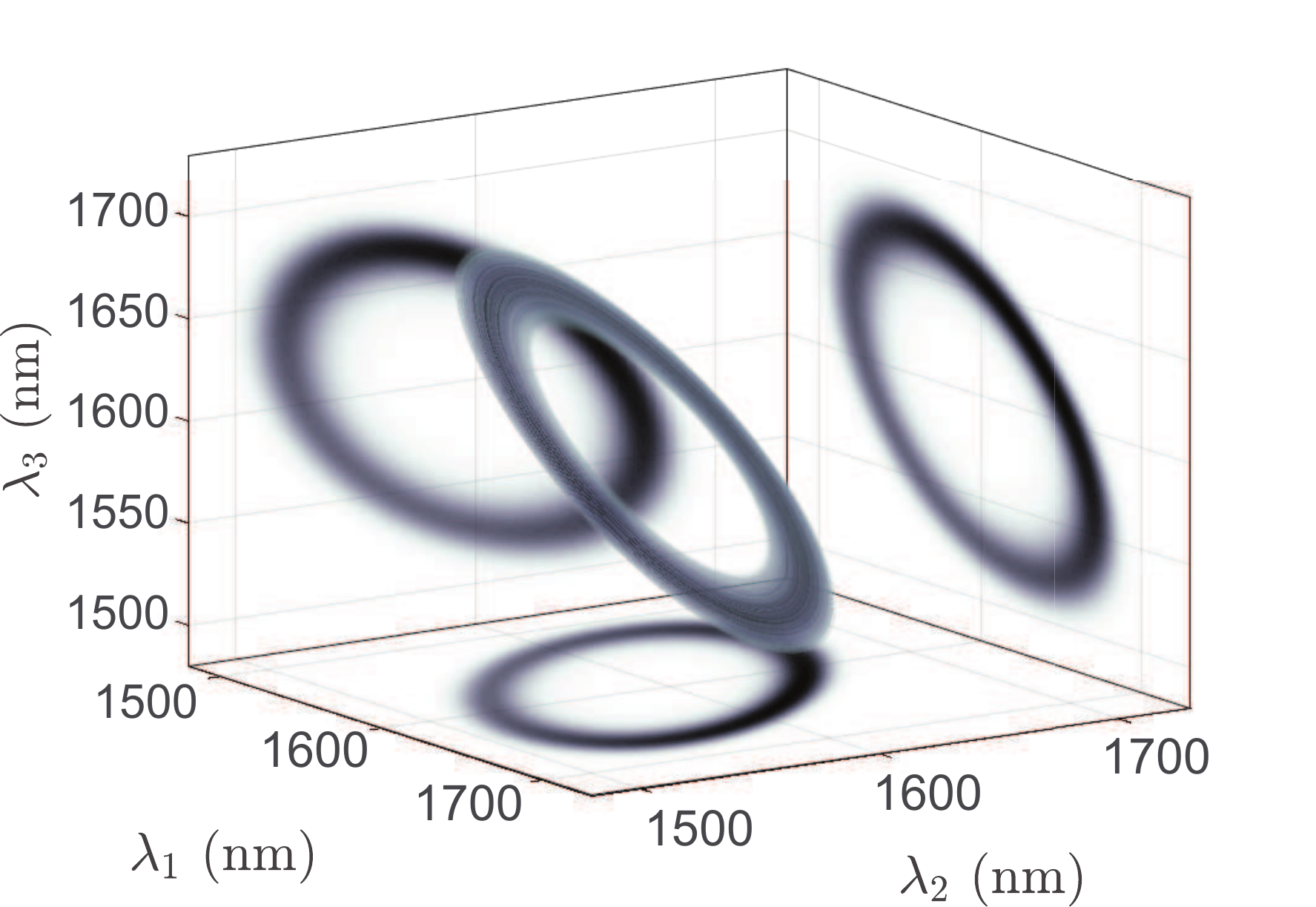}
\caption{Plot of the TOSPDC joint spectral intensity $|f(\lambda_1,\lambda_2,\lambda_3)|^2$ in the frequency non-degenerate source configuration.   Marginal distributions are shown on each of the three coordinate planes.}
\label{Fig:JSITripletesNonDeg1}
\end{figure}

In this section, we present the results of simulations of the expected stimulated throughputs and emission spectra for a specific TOSPDC configuration.   The source characteristics assumed here are the same ones as used in a previous proposal from our group \cite{Corona2011a,Corona2011}, with a non-linear medium in the form of a thin optical fiber with core radius of $r=0.395 \mu$m of length $1$cm. The pump is first assumed to  be centered at  $\omega_p=2\pi c/0.532 \mu$m, with  bandwidth of $4.7/2\pi$ THz, travelling in the HE$_{12}$ spatial mode.   The generated TOSPDC modes are centered at $\omega_i=\omega/3$ in the mode HE$_{11}$.   The resulting  TOSPDC JSI is shown in Fig. \ref{Fig:JSITripletes} , with darker shades of gray  representing higher probabilities of emission. On the walls we have shown plots of the two-photon marginal spectral distributions obtained by tracing over one of the generation modes.   Note that this configuration corresponds to the degenerate case, with triplet emission peaked at $\omega_1=\omega_2=\omega_3=\omega_p/3$.  Note also that the concavity of the JSI is towards the origin, in the joint emission wavelength space, as is evident from the fact that the two-photon marginals extend towards $\lambda < 3 \lambda_p$ (with $\lambda_p= 2 \pi c / \omega_p$). 

In Fig. \ref{Fig:JSITripletesNonDeg1}  we depict the TOSPDC JSI for the experimental situation as above, except that the pump frequency is shifted from $\omega_p=2\pi c/0.532 \mu$m, to $2\pi c/0.531$nm. It is clear from the absence of an emission  maximum at $\omega_p/3$ that this corresponds to a spectrtally  non-degenerate source configuration.    

\begin{figure}
\includegraphics[width=0.45\textwidth]{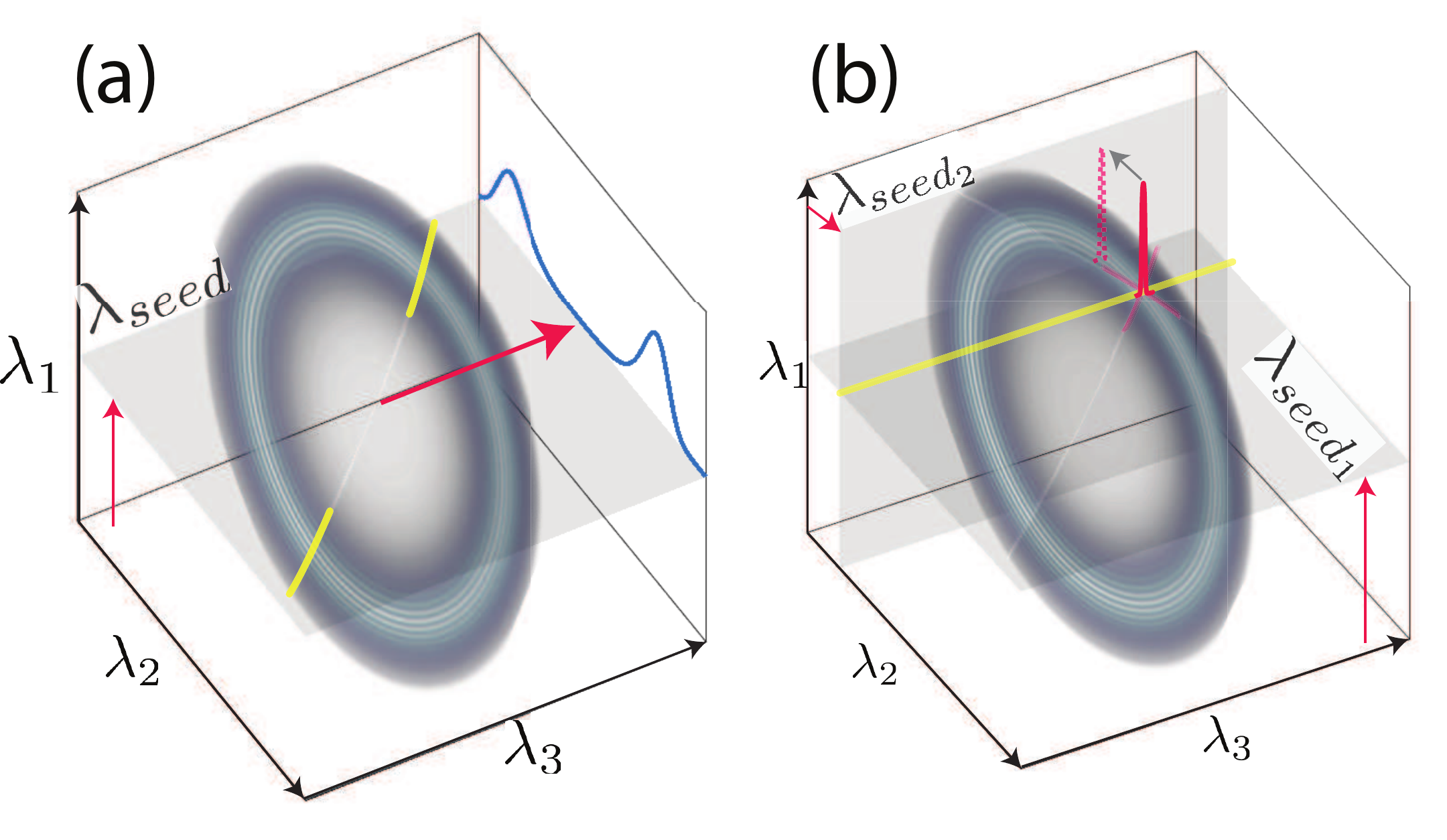}
\caption{Schematic for:  (a)  singly-overlapped seeding, which corresponds to the intersection of a plane at the seed frequency with the joint spectral intensity.  (b) doubly-overlapped seeding, which corresponds to the intersection of two orthogonal planes defined by each of the two seeds with the  joint spectral intensity.  }
\label{Fig:Cartoon}
\end{figure}

Note that in accordance with eq.(\ref{eq:H_I}), and (\ref{Eq:overlap}a), single seeding corresponds to taking a `slice' of the TOSPDC JSI at the seed frequency, i.e. to the intersection between the three-dimensional JSI and a plane placed at the seed frequency, as shown schematically for the non-degenerate case,  in Fig. \ref{Fig:Cartoon}(a).    The function thus obtained, with two frequency arguments, can be either:  i) integrated over both frequency arguments  for the total seeded flux, or ii) integrated over one of the frequency arguments for the seeded emission spectrum (shown as a blue curve in Fig. \ref{Fig:Cartoon}(a).  Similarly, in accordance with  Eq. (\ref{eq:H_II}), and (\ref{Eq:overlap}b),  double seeding corresponds to taking a double slice, i.e. to the triple intersection between the JSI,  a plane placed at the first seed frequency, and a second plane orthogonal to the first placed at the second seed frequency, as shown schematically for the non-degenerate case,  in Fig. \ref{Fig:Cartoon}(b).  The function thus obtained may be i) left intact for the emission spectrum, or ii)  integrated over the frequency argument for the total flux,

\begin{figure}
\includegraphics[width=0.45\textwidth]{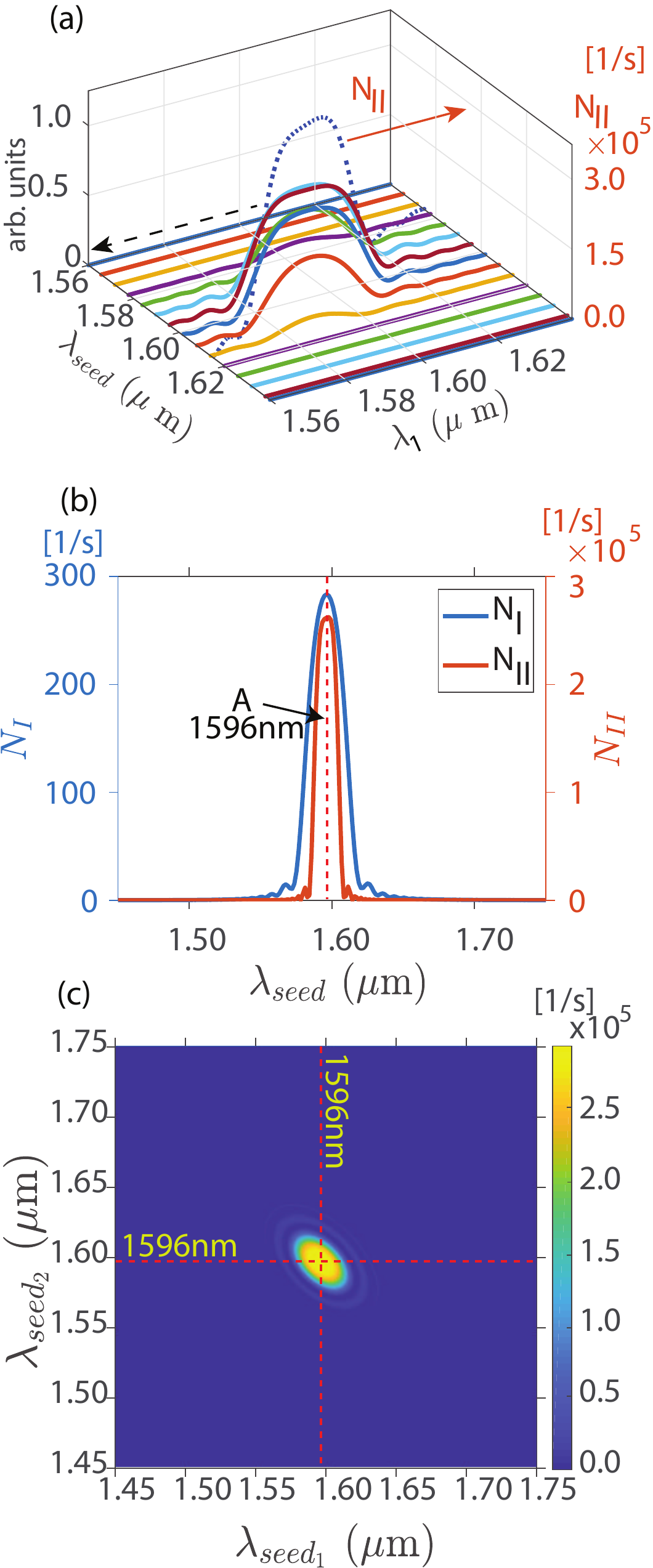}
\caption{For the frequency-degenerate source configuration: (a) Colored continuous lines indicate spectra obtained from singly-overlapped seeding, while the dotted line indicates doubly-overlapped seeding at degenerate seed frequencies.  (b) the blue line is the total flux at each seed frequency obtained as the integral of the spectra in panel (a), while the red line indicates doubly-overlapped seeding at degenerate seed frequencies also shown in (a).    (c) Doubly-overlapped seeded flux obtained with independently-varying seed frequencies.}
\label{Fig:JSITripletesDeg}
\end{figure}

\begin{figure}
\includegraphics[width=0.45\textwidth]{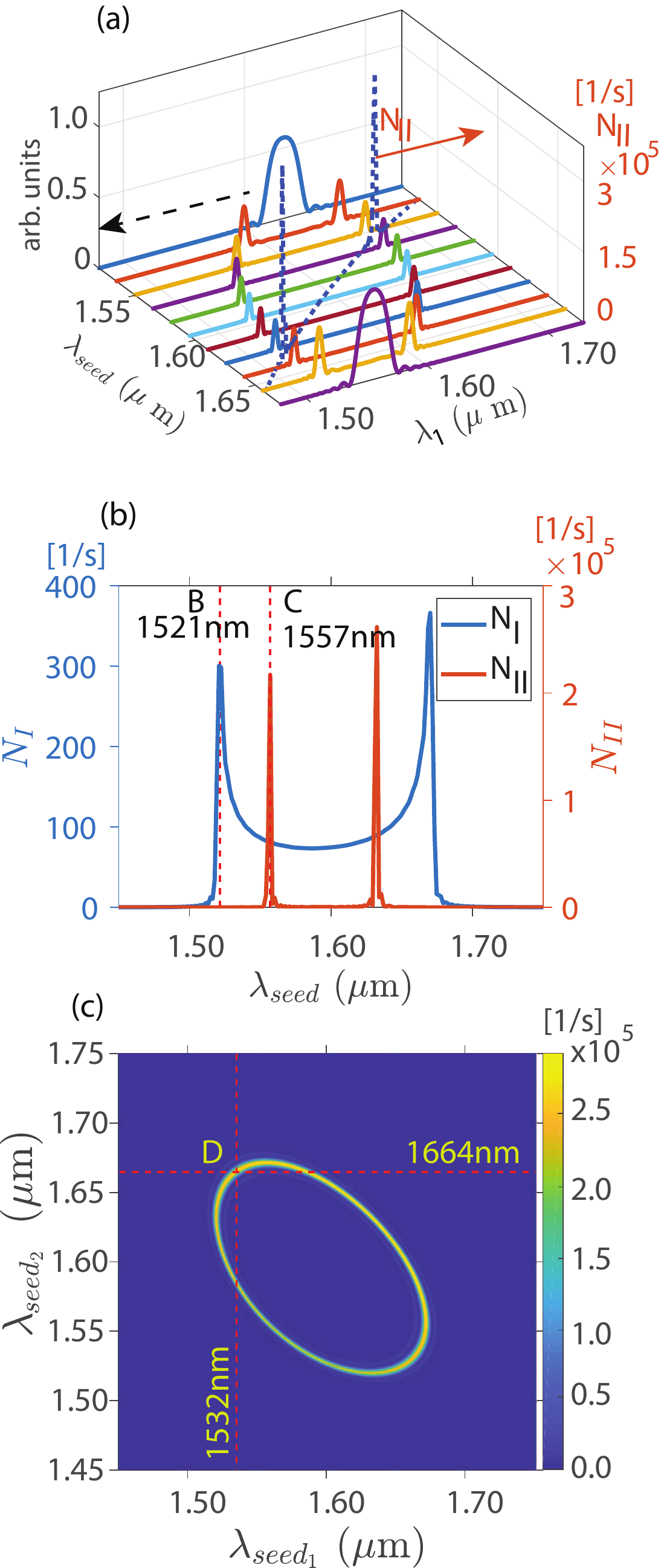}
\caption{For the frequency non-degenerate source configuration: (a) Colored continuous lines indicate spectra obtained from singly-overlapped seeding, while the dotted line indicates doubly-overlapped seeding at degenerate seed frequencies.  (b) the blue line is the total flux at each seed frequency obtained as the integral of the spectra in panel (a), while the red line indicates doubly-overlapped seeding at degenerate seed frequencies also shown in (a).    (c) Doubly-overlapped seeded flux obtained with independently-varying seed frequencies. }
\label{Fig:JSITripletesNonDeg}
\end{figure}

We now proceed to present numerical evaluations of the seeded throughputs obtained for various situations of interest, based on the source described above.     We are  interested in comparing the behavior, in the presence of seeding,  of the degenerate and non-degenerate source configurations (see Figures \ref{Fig:JSITripletesDeg} and \ref{Fig:JSITripletesNonDeg}).    For both of these configurations, we are also interested in comparing the resulting behavior when the pump and seed fields 
are selected as pulsed or monochromatic in all possible combinations.   

At first we assume that both the pump and seed are monochromatic.  In order to  be able to provide numerical estimates for the throughputs, we (arbitrarily) assume a pump power of $200$mW and a seed power of $10$mW.  For the degenerate source configuration resulting from a pump wavelength $532$nm, and in the presence of a single seed frequency,  Fig. \ref{Fig:JSITripletesDeg}(a) shows the emitted spectra (colored solid lines) $N_1(\lambda_r)$  obtained for a number of different seed frequencies $\lambda_{seed}$, as derived from singly-overlapped seeding.   The dashed line shows the doubly-overlapped (i.e. frequency-degenerate double seed) throughput obtained in the presence of a single seed frequency $\omega_{seed}$ at frequencies $\omega_r$ which fulfil the energy conservation constraint $\omega_r= 2 \omega_p-\omega_{seed}$.

Fig. \ref{Fig:JSITripletesDeg}(b) shows (blue curve) the total flux expected for each seed frequency, obtained by integrating the individual singly-seeded spectra from panel (a).    So as to compare with the degenerate double-seed case, we also present in the same axis the doubly-seeded behavior (red curve), already shown in panel (a).     It becomes clear from this figure that the doubly-seeded case leads to three orders of magnitude greater flux as compared to the singly-seeed cases.   Note that the doubly-seeded flux is indeed expected to be greater than the singly-seeded flux by a factor proportional to the product of the second seed intensity and the quotient of the overlap terms  $\Theta_2/\Theta_1$ (see Eqs. (\ref{eq:N1N2STOPDC}) and (\ref{eq:N1N2comp})), which in this case amounts to these three orders of magnitude.    Finally, panel (c) shows the doubly-seeded throughput obtained by letting the two seed frequencies $\omega_{seed1}$ and $\omega_{seed2}$ vary independently.    Note that we can recover the red curve in panel (b) by evaluating this non-degenerate doubly-seed response along the line $\omega_{seed1}=\omega_{seed2}$. 
  
Let us now turn our attention to the non-degenerate source configuration, with a pump wavelength $\lambda_p=531$nm, for the case where both the pump and seed are monochromatic. Again, we assume a pump power of $200$mW and a seed power of $10$mW.   Fig. \ref{Fig:JSITripletesNonDeg}(a) shows the emitted spectra (colored solid lines) $N_1(\lambda_r)$  obtained for a number of different seed frequencies $\lambda_{seed}$, as derived from singly-overlapped seeding.   The dashed line shows the doubly-overlapped (i.e. frequency-degenerate double seed) throughput obtained in the presence of a single seed frequency $\omega_{seed}$ at frequencies $\omega_r$ which fulfil the energy conservation constraint $\omega_r= 2 \omega_p-\omega_{seed}$.

Fig. \ref{Fig:JSITripletesNonDeg}(b) shows (blue curve) the total flux expected for each seed frequency, obtained by integrating the individual singly-seeded spectra from panel (a).    So as to compare with the degenerate double-seed case, we also present in the same axis the doubly-seeded behavior (red curve), already shown in panel (a).     As for the degenerate source configuration,  the doubly-seeded case leads to three orders of magnitude greater flux as compared to the singly-seeed cases.  Also note that in contrast with the degenerate source configuration, the frequency-degenerate doubly-seeded case is in the form of two sharp peaks while the singly-seeded contribution is spectrally broad.   Finally, panel (c) shows the doubly-seeded throughput obtained by letting the two seed frequencies $\omega_{seed1}$ and $\omega_{seed2}$ vary independently.    Note that we  can recover the red curve in panel (b) by evaluating this non-degenerate doubly-seed response along the line $\omega_{seed1}=\omega_{seed2}$.   Also note that in contrast with the degenerate source configuration this $N_2$ behavior with non-degenerate arguments is in the form of a ring instead of a single broad peak.

We have contrasted the behavior, under singly- and doubly- overlapped seeding, of the degenerate and non-degenerate source configurations.   In order to compare for each of these configurations the behavior when each of the pump and seed are allowed to be pulsed or monochromatic, we select four spectral points from Figs. \ref{Fig:JSITripletesDeg} and \ref{Fig:JSITripletesNonDeg}, and show the resulting throughputs in Table \ref{Table01}.  Point A, with $\lambda_1=1596$nm,  corresponds to the location of the maximum rate of seeded throughput for the degenerate source configuration.   In the presence of a single seed wavelength for the non-degenerate source configuration, Point B with $\lambda_1=1521$nm,  corresponds to one of two maxima of the singly-overlapped seeded throughput, while point C with  $\lambda_1=1557$nm, corresponds to one of two maxima of the doubly-overlapped seeded throughgput.  Finally, point D with $\lambda_1=1532$nm  and $\lambda_2=1664$nm, corresponds to a non-degenerate selection of seeds, both exhibiting overlap with the JSI. 

Note, for point A, that while the throughput difference  between the doubly-overlapped and singly-overlapped cases is 3 orders of magnitude for the monochromatic-monochromatic situation (as was already pointed out), this difference grows to a remarkable 8 orders of magnitude for the pulsed-pulsed situation.   Points B and C illustrate that at a singly-overlapped (and non-doubly-overlapped) spectral location (i.e. point B), $N_2$  drops sharply as expected, compared to point C where both types of overlap occur.   Nevertheless, the drop in $N_2$ for point  B is orders of magnitude less severe for the pulsed-pulsed situation, as compared to the monochromatic-monochromatic situation, since for the former the non-zero bandwidths involved ensure that some overlap with the JSI survives.  Point D illustrates that for double-overlap with dissimilar seed frequencies, the resulting throughput is similar as compared to the case of frequency-degenerate seeds. 

In obtaining the values shown in the table, we have assumed for the pulsed configurations a pump bandwidth of $\sigma_p=\num{4.7e12}/2\pi$ Hz  and a seed bandwidth  $\sigma_s$ of a tenth of this value, i.e.  $\sigma_s=\sigma_p/10$, while we have assumed a repetition rate of $10$MHz (for both pump and seed).  In the case where both fields are pulsed we have assumed that they are perfectly temporally matched.
It is clear from these results that the pulsed-pulsed situation leads to the greatest emitted flux, 4 (9) orders of magnitude larger as compared to that obtained in the monochromatic-monochromatic situation for singly-overlapped (doubly-overlapped) seeding.  The mixed cases, i.e. monochromatic-pulsed and pulsed-monochromartic, are clearly less interesting with a much reduced flux due to the resulting hampered temporal matching between pump and seed.  Obtaining perfect temporal matching between a pulsed pump and a pulsed seed may be challenging in practice unless  one of them gives rise to the other through an appropriate  non-linear process \cite{Agafonov2010}.   In cases where such pulsed temporal matching is unfeasible, the monochromatic-monochromatic situation is claerly the best alternative.

\begin{table}
\caption{Comparison of the resulting seeded throughput for pump and seed in different combinations of being pulsed and monochromatic (MC), at spectral points A (pertaining to the degenerate, $\mathscr{D}$, source configuration), and points B,C, and D  (pertaining to the non-degenerate, $\mathscr{ND}$, source configuration), as indicated in Figs. \ref{Fig:JSITripletesDeg} and   \ref{Fig:JSITripletesNonDeg}. }
\resizebox{0.5\textwidth}{!} {%
\begin{tabular}{llclll}\label{Table01}
& & &Wavelength & $N_I (\lambda)$ [photons$\cdot s^{-1}]$  &$N_{II} (\lambda_{s_1},\lambda_{s_2} )$ [photons$\cdot s^{-1}]$ \\ 
\hline
\hline
\parbox[t]{2mm}{\multirow{6}{*}{\rotatebox[origin=c]{90}{Pulsed-Pulsed}}}&\parbox[t]{2mm}{\multirow{1}{*}{\rotatebox[origin=c]{90}{$\mathscr{D}$}}} &
A & $\lambda_1 = 1596nm$ & $N_I (\lambda_1) = \num{4.0e6}$ & $N_{II} (\lambda_1,\lambda_1) = \num{1.025e14}$   \\ 
\cline{2-6}
&\parbox[t]{2mm}{\multirow{5}{*}{\rotatebox[origin=c]{90}{$\mathscr{ND}$}}} &
B& $\lambda_1 = 1521nm$ & $N_I (\lambda_1) = \num{4.0e6}$ & $N_{II} (\lambda_1,\lambda_1) = \num{1.1e11}  $ \\ 
\cline{3-6}
&&C & $\lambda_1 =1557nm$ & $N_I (\lambda_1) = \num{3.8e6} $ & $N_{II} (\lambda_1,\lambda_1) = \num{9.8e13}   $ \\ 
\cline{3-6} 
&&D& $\lambda_1 = 1532nm$ & $N_I (\lambda_1) =\num{4.8e6} $  & $N_{II} (\lambda_1,\lambda_1) = \num{3.6e11}   $ \\ 
&& & $\lambda_2 = 1664nm$ & $N_I (\lambda_2) = \num{4.4e6}$  & $N_{II} (\lambda_2,\lambda_2)=\num{1.3e10}$  \\ 
&& & & & $N_{II} (\lambda_1,\lambda_2) = \num{1.0e14}$ \\
\hline 
\hline
\parbox[t]{2mm}{\multirow{6}{*}{\rotatebox[origin=c]{90}{MC-MC}}}&\parbox[t]{2mm}{\multirow{1}{*}{\rotatebox[origin=c]{90}{$\mathscr{D}$}}} &
A & $\lambda_1 = 1596nm$ & $N_I (\lambda_1) =\num{2.8e2}$ & $N_{II} (\lambda_1,\lambda_1) = \num{2.6e5}$   \\ 
\cline{2-6}
&\parbox[t]{2mm}{\multirow{5}{*}{\rotatebox[origin=c]{90}{$\mathscr{ND}$}}} &
B& $\lambda_1 = 1521nm$ & $N_I (\lambda_1) = \num{3.0e2}$ & $N_{II} (\lambda_1,\lambda_1) = 14 $ \\ 
\cline{3-6}
&&C & $\lambda_1 =1557nm$ & $N_I (\lambda_1) = 82 $ & $N_{II} (\lambda_1,\lambda_1) = \num{2.2e5}   $ \\ 
\cline{3-6}
&&D& $\lambda_1 = 1532nm$ & $N_I (\lambda_1) =\num{1.25e2} $  &  $N_{II} (\lambda_1,\lambda_1) = \num{2.6}   $ \\ 
&& & $\lambda_2 = 1664nm$ & $N_I (\lambda_2) = \num{2.0e2}$  &  $N_{II} (\lambda_2,\lambda_2) = \num{88}   $ \\ 
&&& & & $N_{II} (\lambda_1,\lambda_2) = \num{2.5e5}   $ \\

\hline
\hline
\parbox[t]{2mm}{\multirow{6}{*}{\rotatebox[origin=c]{90}{Pulsed-MC}}}&\parbox[t]{2mm}{\multirow{1}{*}{\rotatebox[origin=c]{90}{$\mathscr{D}$}}} & A & $\lambda_1 = 1596nm$ & $N_I (\lambda_1) = \num { 9.7e-12}$ & $N_{II} (\lambda_1,\lambda_1) = \num{1.4e-9}$   \\ 
\cline{2-6}
&\parbox[t]{2mm}{\multirow{5}{*}{\rotatebox[origin=c]{90}{$\mathscr{ND}$}}} & B& $\lambda_1 = 1521nm$ & $N_I (\lambda_1) = \num{8.9e-12}$ & $N_{II} (\lambda_1,\lambda_1) = \num{4.8e-13}  $ \\ 
\cline{3-6}
&&C & $\lambda_1 =1557nm$ & $N_I (\lambda_1) = \num{8.418e-12} $ & $N_{II} (\lambda_1,\lambda_1) = \num{1.1e-9}   $ \\
\cline{3-6}
&&D& $\lambda_1 = 1532nm$ & $N_I (\lambda_1) =\num{1.1e-11} $  & $N_{II} (\lambda_1,\lambda_1) = \num{2.3e-12}   $ \\ 
&& & $\lambda_2 = 1664nm$ & $N_I (\lambda_2) = \num{1.3e-11}$  & $N_{II} (\lambda_2,\lambda_2)=\num{1.8e-12}$  \\ 
 &&& & & $N_{II} (\lambda_1,\lambda_2) = \num{1.4e-9}$ \\
\hline 
\hline
\parbox[t]{2mm}{\multirow{6}{*}{\rotatebox[origin=c]{90}{MC-Pulsed}}}&\parbox[t]{2mm}{\multirow{1}{*}{\rotatebox[origin=c]{90}{$\mathscr{D}$}}} & A & $\lambda_1 = 1596nm$ & $N_I (\lambda_1) = \num {3.8e-10}$ & $N_{II} (\lambda_1,\lambda_1) = \num{6.0e-3}$   \\ 
\cline{2-6}
&\parbox[t]{2mm}{\multirow{5}{*}{\rotatebox[origin=c]{90}{$\mathscr{ND}$}}} & B& $\lambda_1 = 1521nm$ & $N_I (\lambda_1) = \num{4.0e-10}$ & $N_{II} (\lambda_1,\lambda_1) = \num{0}  $ \\ 
\cline{3-6}
&& C & $\lambda_1 =1557nm$ & $N_I (\lambda_1) = \num{1.1e-10} $ & $N_{II} (\lambda_1,\lambda_1) = \num{5.7e-3}   $ \\ 
\cline{3-6}
&& D& $\lambda_1 = 1532nm$ & $N_I (\lambda_1) =\num{1.7e-10} $  & $N_{II} (\lambda_1,\lambda_1) = \num{3.0e-7}   $ \\ 
& & & $\lambda_2 = 1664nm$ & $N_I (\lambda_2) = \num{2.6e-10}$  & $N_{II} (\lambda_2,\lambda_2)=\num{3.7e-27}$  \\ 
& & & & & $N_{II} (\lambda_1,\lambda_2) = \num{5.2e-3}$ \\
\hline 
\end{tabular} 
}
\end{table}

\section{Conclusions}

In conclusion, we have analyzed theoretically as well as numerically,  the process of  stimulated third-order parametric downconversion (STOPDC). The work is based on our previous studies of third-order \emph{spontaneous} parametric dowcnonversion (TOSPDC), with the addition of seeding.   We present a calculation leading to expressions for the seeded throughput, which is a direct generalization of previously-reported studies \cite{Liscidini2013}  on second-order stimulated  parametric downconversion.    In our analysis, we allow the seed or seeds to overlap  one or two of the TOSDPC modes, and likewise we allow the pump and seed fields to be either monochromatic or pulsed.    We present general expressions for the spectra and flux produced by the STOPDC process, as well as a numerical study for a particular source design.    We conclude from our numerical study that doubly-overlapped seeding can lead to a considerably greater flux (in the cases shown by up to eight orders of magnitude) as compared to singly-overlapped seeding.    We furthermore describe how doubly-overlapped seeding may be employed as the  basis for stimulated emission tomography which allows for the reconstruction of the three-photon joint spectral amplitude. We find that among the different combinations of monochromatic and pulsed nature for the pump and seed fields, the pulsed-pulsed and monochromatic-monochromatic cases lead to much greater throughputs as compared to the mixed pulsed-monochromatic cases.  While the pulsed-pulsed situation is the ideal one permitting the greatest seeded flux, the difficulty of attaining temporal matching between two independent pulse trains makes the monochromatic-monochromatic an attractive alternative.    We hope that this work will guide future experimental work on seeded third-order parametric downconversion.

\begin{acknowledgements}
FD acknowledges support from National Council of Science and Technology of Mexico - CONACYT (C\'atedras CONACYT 709/2018); AU from PAPIIT (UNAM) grant IN104418, CONACYT Fronteras de la Ciencia grant 1667, and AFOSR grant FA9550-16-1-1458.	
	
\end{acknowledgements}


\begin{thebibliography}{42}%
	\makeatletter
	\providecommand \@ifxundefined [1]{%
		\@ifx{#1\undefined}
	}%
	\providecommand \@ifnum [1]{%
		\ifnum #1\expandafter \@firstoftwo
		\else \expandafter \@secondoftwo
		\fi
	}%
	\providecommand \@ifx [1]{%
		\ifx #1\expandafter \@firstoftwo
		\else \expandafter \@secondoftwo
		\fi
	}%
	\providecommand \natexlab [1]{#1}%
	\providecommand \enquote  [1]{``#1''}%
	\providecommand \bibnamefont  [1]{#1}%
	\providecommand \bibfnamefont [1]{#1}%
	\providecommand \citenamefont [1]{#1}%
	\providecommand \href@noop [0]{\@secondoftwo}%
	\providecommand \href [0]{\begingroup \@sanitize@url \@href}%
	\providecommand \@href[1]{\@@startlink{#1}\@@href}%
	\providecommand \@@href[1]{\endgroup#1\@@endlink}%
	\providecommand \@sanitize@url [0]{\catcode `\\12\catcode `\$12\catcode
		`\&12\catcode `\#12\catcode `\^12\catcode `\_12\catcode `\%12\relax}%
	\providecommand \@@startlink[1]{}%
	\providecommand \@@endlink[0]{}%
	\providecommand \url  [0]{\begingroup\@sanitize@url \@url }%
	\providecommand \@url [1]{\endgroup\@href {#1}{\urlprefix }}%
	\providecommand \urlprefix  [0]{URL }%
	\providecommand \Eprint [0]{\href }%
	\providecommand \doibase [0]{http://dx.doi.org/}%
	\providecommand \selectlanguage [0]{\@gobble}%
	\providecommand \bibinfo  [0]{\@secondoftwo}%
	\providecommand \bibfield  [0]{\@secondoftwo}%
	\providecommand \translation [1]{[#1]}%
	\providecommand \BibitemOpen [0]{}%
	\providecommand \bibitemStop [0]{}%
	\providecommand \bibitemNoStop [0]{.\EOS\space}%
	\providecommand \EOS [0]{\spacefactor3000\relax}%
	\providecommand \BibitemShut  [1]{\csname bibitem#1\endcsname}%
	\let\auto@bib@innerbib\@empty
	\bibitem [{\citenamefont {Bennett}\ and\ \citenamefont
		{DiVincenzo}(2000)}]{Bennett2000}%
	\BibitemOpen
	\bibfield  {author} {\bibinfo {author} {\bibfnamefont {C.~H.}\ \bibnamefont
			{Bennett}}\ and\ \bibinfo {author} {\bibfnamefont {D.~P.}\ \bibnamefont
			{DiVincenzo}},\ }\href {\doibase 10.1063/1.881452} {\bibfield  {journal}
		{\bibinfo  {journal} {Nature}\ }\textbf {\bibinfo {volume} {404}},\ \bibinfo
		{pages} {247} (\bibinfo {year} {2000})}\BibitemShut {NoStop}%
	\bibitem [{\citenamefont {Jozsa}\ and\ \citenamefont
		{Linden}(2003)}]{Jozsa2003}%
	\BibitemOpen
	\bibfield  {author} {\bibinfo {author} {\bibfnamefont {R.}~\bibnamefont
			{Jozsa}}\ and\ \bibinfo {author} {\bibfnamefont {N.}~\bibnamefont {Linden}},\
	}\href {\doibase 10.1098/rspa.2002.1097} {\bibfield  {journal} {\bibinfo
			{journal} {Proceedings of the Royal Society A: Mathematical, Physical and
				Engineering Sciences}\ }\textbf {\bibinfo {volume} {459}},\ \bibinfo {pages}
		{2011} (\bibinfo {year} {2003})}\BibitemShut {NoStop}%
	\bibitem [{\citenamefont {Steane}(1998)}]{Hughes2007}%
	\BibitemOpen
	\bibfield  {author} {\bibinfo {author} {\bibfnamefont {A.}~\bibnamefont
			{Steane}},\ }\href {\doibase 10.1088/0034-4885/61/2/002} {\bibfield
		{journal} {\bibinfo  {journal} {Rep. Prog. Phys.}\ }\textbf {\bibinfo
			{volume} {61}},\ \bibinfo {pages} {117} (\bibinfo {year} {1998})}\BibitemShut
	{NoStop}%
	\bibitem [{\citenamefont {Prevedel}\ \emph {et~al.}(2007)\citenamefont
		{Prevedel}, \citenamefont {Aspelmeyer}, \citenamefont {Brukner},
		\citenamefont {Zeilinger},\ and\ \citenamefont {Jennewein}}]{Prevedel2007}%
	\BibitemOpen
	\bibfield  {author} {\bibinfo {author} {\bibfnamefont {R.}~\bibnamefont
			{Prevedel}}, \bibinfo {author} {\bibfnamefont {M.}~\bibnamefont
			{Aspelmeyer}}, \bibinfo {author} {\bibfnamefont {C.}~\bibnamefont {Brukner}},
		\bibinfo {author} {\bibfnamefont {A.}~\bibnamefont {Zeilinger}}, \ and\
		\bibinfo {author} {\bibfnamefont {T.~D.}\ \bibnamefont {Jennewein}},\ }\href
	{\doibase 10.1364/josab.24.000241} {\bibfield  {journal} {\bibinfo  {journal}
			{Journal of the Optical Society of America B}\ }\textbf {\bibinfo {volume}
			{24}},\ \bibinfo {pages} {241} (\bibinfo {year} {2007})}\BibitemShut
	{NoStop}%
	\bibitem [{\citenamefont {Vedral}(2014)}]{Vedral2014}%
	\BibitemOpen
	\bibfield  {author} {\bibinfo {author} {\bibfnamefont {V.}~\bibnamefont
			{Vedral}},\ }\href {\doibase 10.1038/nphys2904} {\bibfield  {journal}
		{\bibinfo  {journal} {Nature Physics}\ }\textbf {\bibinfo {volume} {10}},\
		\bibinfo {pages} {256} (\bibinfo {year} {2014})}\BibitemShut {NoStop}%
	\bibitem [{\citenamefont {Brecht}\ \emph {et~al.}(2015)\citenamefont {Brecht},
		\citenamefont {Reddy}, \citenamefont {Silberhorn},\ and\ \citenamefont
		{Raymer}}]{Brecht2015}%
	\BibitemOpen
	\bibfield  {author} {\bibinfo {author} {\bibfnamefont {B.}~\bibnamefont
			{Brecht}}, \bibinfo {author} {\bibfnamefont {D.~V.}\ \bibnamefont {Reddy}},
		\bibinfo {author} {\bibfnamefont {C.}~\bibnamefont {Silberhorn}}, \ and\
		\bibinfo {author} {\bibfnamefont {M.~G.}\ \bibnamefont {Raymer}},\ }\href
	{\doibase 10.1103/PhysRevX.5.041017} {\bibfield  {journal} {\bibinfo
			{journal} {Physical Review X}\ }\textbf {\bibinfo {volume} {5}},\ \bibinfo
		{pages} {041017} (\bibinfo {year} {2015})}\BibitemShut {NoStop}%
	\bibitem [{\citenamefont {O'Brien}\ \emph {et~al.}(2009)\citenamefont
		{O'Brien}, \citenamefont {Furusawa},\ and\ \citenamefont
		{Vu{\v{c}}kovi{\'{c}}}}]{Obrien2010}%
	\BibitemOpen
	\bibfield  {author} {\bibinfo {author} {\bibfnamefont {J.~L.}\ \bibnamefont
			{O'Brien}}, \bibinfo {author} {\bibfnamefont {A.}~\bibnamefont {Furusawa}}, \
		and\ \bibinfo {author} {\bibfnamefont {J.}~\bibnamefont
			{Vu{\v{c}}kovi{\'{c}}}},\ }\href {\doibase 10.1038/nphoton.2009.229}
	{\bibfield  {journal} {\bibinfo  {journal} {Nature Photonics}\ }\textbf
		{\bibinfo {volume} {3}},\ \bibinfo {pages} {687} (\bibinfo {year}
		{2009})}\BibitemShut {NoStop}%
	\bibitem [{\citenamefont {Knill}\ \emph {et~al.}(2001)\citenamefont {Knill},
		\citenamefont {Laflamme},\ and\ \citenamefont {Milburn}}]{Knill01a}%
	\BibitemOpen
	\bibfield  {author} {\bibinfo {author} {\bibfnamefont {E.}~\bibnamefont
			{Knill}}, \bibinfo {author} {\bibfnamefont {R.}~\bibnamefont {Laflamme}}, \
		and\ \bibinfo {author} {\bibfnamefont {G.~J.}\ \bibnamefont {Milburn}},\
	}\href@noop {} {\bibfield  {journal} {\bibinfo  {journal} {Nature}\ }\textbf
		{\bibinfo {volume} {409}},\ \bibinfo {pages} {46} (\bibinfo {year}
		{2001})}\BibitemShut {NoStop}%
	\bibitem [{\citenamefont {O'Brien}(2007)}]{OBrien2007}%
	\BibitemOpen
	\bibfield  {author} {\bibinfo {author} {\bibfnamefont {J.~L.}\ \bibnamefont
			{O'Brien}},\ }\href {\doibase 10.1126/science.1142892} {\bibfield  {journal}
		{\bibinfo  {journal} {Science}\ }\textbf {\bibinfo {volume} {318}},\ \bibinfo
		{pages} {1567} (\bibinfo {year} {2007})}\BibitemShut {NoStop}%
	\bibitem [{\citenamefont {Burnham}\ and\ \citenamefont
		{Weinberg}(1970)}]{Burnham1970}%
	\BibitemOpen
	\bibfield  {author} {\bibinfo {author} {\bibfnamefont {D.~C.}\ \bibnamefont
			{Burnham}}\ and\ \bibinfo {author} {\bibfnamefont {D.~L.}\ \bibnamefont
			{Weinberg}},\ }\href {\doibase 10.1103/PhysRevLett.25.84} {\bibfield
		{journal} {\bibinfo  {journal} {Physical Review Letters}\ }\textbf {\bibinfo
			{volume} {25}},\ \bibinfo {pages} {84} (\bibinfo {year} {1970})}\BibitemShut
	{NoStop}%
	\bibitem [{\citenamefont {Cohen}\ \emph {et~al.}(2009)\citenamefont {Cohen},
		\citenamefont {Lundeen}, \citenamefont {Smith}, \citenamefont {Puentes},
		\citenamefont {Mosley},\ and\ \citenamefont {Walmsley}}]{Cohen2009}%
	\BibitemOpen
	\bibfield  {author} {\bibinfo {author} {\bibfnamefont {O.}~\bibnamefont
			{Cohen}}, \bibinfo {author} {\bibfnamefont {J.~S.}\ \bibnamefont {Lundeen}},
		\bibinfo {author} {\bibfnamefont {B.~J.}\ \bibnamefont {Smith}}, \bibinfo
		{author} {\bibfnamefont {G.}~\bibnamefont {Puentes}}, \bibinfo {author}
		{\bibfnamefont {P.~J.}\ \bibnamefont {Mosley}}, \ and\ \bibinfo {author}
		{\bibfnamefont {I.~A.}\ \bibnamefont {Walmsley}},\ }\href {\doibase
		10.1103/PhysRevLett.102.123603} {\bibfield  {journal} {\bibinfo  {journal}
			{Physical Review Letters}\ }\textbf {\bibinfo {volume} {102}},\ \bibinfo
		{pages} {123603} (\bibinfo {year} {2009})}\BibitemShut {NoStop}%
	\bibitem [{\citenamefont {Pittman}\ \emph {et~al.}(2005)\citenamefont
		{Pittman}, \citenamefont {Jacobs},\ and\ \citenamefont
		{Franson}}]{Pittman2005}%
	\BibitemOpen
	\bibfield  {author} {\bibinfo {author} {\bibfnamefont {T.~B.}\ \bibnamefont
			{Pittman}}, \bibinfo {author} {\bibfnamefont {B.~C.}\ \bibnamefont {Jacobs}},
		\ and\ \bibinfo {author} {\bibfnamefont {J.~D.}\ \bibnamefont {Franson}},\
	}\href {\doibase 10.1016/j.optcom.2004.11.027} {\bibfield  {journal}
		{\bibinfo  {journal} {Optics Communications}\ }\textbf {\bibinfo {volume}
			{246}},\ \bibinfo {pages} {545} (\bibinfo {year} {2005})}\BibitemShut
	{NoStop}%
	\bibitem [{\citenamefont {Huang}\ \emph {et~al.}(2011)\citenamefont {Huang},
		\citenamefont {Altepeter},\ and\ \citenamefont {Kumar}}]{Huang2011}%
	\BibitemOpen
	\bibfield  {author} {\bibinfo {author} {\bibfnamefont {Y.~P.}\ \bibnamefont
			{Huang}}, \bibinfo {author} {\bibfnamefont {J.~B.}\ \bibnamefont
			{Altepeter}}, \ and\ \bibinfo {author} {\bibfnamefont {P.}~\bibnamefont
			{Kumar}},\ }\href {\doibase 10.1103/PhysRevA.84.033844} {\bibfield  {journal}
		{\bibinfo  {journal} {Physical Review A - Atomic, Molecular, and Optical
				Physics}\ }\textbf {\bibinfo {volume} {84}},\ \bibinfo {pages} {033844}
		(\bibinfo {year} {2011})}\BibitemShut {NoStop}%
	\bibitem [{\citenamefont {Meyer-Scott}\ \emph {et~al.}(2017)\citenamefont
		{Meyer-Scott}, \citenamefont {Montaut}, \citenamefont {Tiedau}, \citenamefont
		{Sansoni}, \citenamefont {Herrmann}, \citenamefont {Bartley},\ and\
		\citenamefont {Silberhorn}}]{Meyer-Scott2017}%
	\BibitemOpen
	\bibfield  {author} {\bibinfo {author} {\bibfnamefont {E.}~\bibnamefont
			{Meyer-Scott}}, \bibinfo {author} {\bibfnamefont {N.}~\bibnamefont
			{Montaut}}, \bibinfo {author} {\bibfnamefont {J.}~\bibnamefont {Tiedau}},
		\bibinfo {author} {\bibfnamefont {L.}~\bibnamefont {Sansoni}}, \bibinfo
		{author} {\bibfnamefont {H.}~\bibnamefont {Herrmann}}, \bibinfo {author}
		{\bibfnamefont {T.~J.}\ \bibnamefont {Bartley}}, \ and\ \bibinfo {author}
		{\bibfnamefont {C.}~\bibnamefont {Silberhorn}},\ }\href {\doibase
		10.1103/PhysRevA.95.061803} {\bibfield  {journal} {\bibinfo  {journal}
			{Physical Review A}\ }\textbf {\bibinfo {volume} {95}},\ \bibinfo {pages}
		{061803} (\bibinfo {year} {2017})}\BibitemShut {NoStop}%
	\bibitem [{\citenamefont {Zhang}\ \emph {et~al.}(2012)\citenamefont {Zhang},
		\citenamefont {S{\"{o}}ller}, \citenamefont {Cohen}, \citenamefont {Smith},\
		and\ \citenamefont {Walmsley}}]{Zhang2012}%
	\BibitemOpen
	\bibfield  {author} {\bibinfo {author} {\bibfnamefont {L.}~\bibnamefont
			{Zhang}}, \bibinfo {author} {\bibfnamefont {C.}~\bibnamefont {S{\"{o}}ller}},
		\bibinfo {author} {\bibfnamefont {O.}~\bibnamefont {Cohen}}, \bibinfo
		{author} {\bibfnamefont {B.~J.}\ \bibnamefont {Smith}}, \ and\ \bibinfo
		{author} {\bibfnamefont {I.~A.}\ \bibnamefont {Walmsley}},\ }\href {\doibase
		10.1080/09500340.2012.679707} {\bibfield  {journal} {\bibinfo  {journal}
			{Journal of Modern Optics}\ }\textbf {\bibinfo {volume} {59}},\ \bibinfo
		{pages} {1525} (\bibinfo {year} {2012})}\BibitemShut {NoStop}%
	\bibitem [{\citenamefont {{\'{S}}liwa}\ and\ \citenamefont
		{Banaszek}(2003)}]{Sliwa2003}%
	\BibitemOpen
	\bibfield  {author} {\bibinfo {author} {\bibfnamefont {C.}~\bibnamefont
			{{\'{S}}liwa}}\ and\ \bibinfo {author} {\bibfnamefont {K.}~\bibnamefont
			{Banaszek}},\ }\href {\doibase 10.1103/PhysRevA.67.030101} {\bibfield
		{journal} {\bibinfo  {journal} {Physical Review A - Atomic, Molecular, and
				Optical Physics}\ }\textbf {\bibinfo {volume} {67}},\ \bibinfo {pages} {4}
		(\bibinfo {year} {2003})}\BibitemShut {NoStop}%
	\bibitem [{\citenamefont {Walther}\ \emph {et~al.}(2007)\citenamefont
		{Walther}, \citenamefont {Aspelmeyer},\ and\ \citenamefont
		{Zeilinger}}]{Walther2007}%
	\BibitemOpen
	\bibfield  {author} {\bibinfo {author} {\bibfnamefont {P.}~\bibnamefont
			{Walther}}, \bibinfo {author} {\bibfnamefont {M.}~\bibnamefont {Aspelmeyer}},
		\ and\ \bibinfo {author} {\bibfnamefont {A.}~\bibnamefont {Zeilinger}},\
	}\href {\doibase 10.1103/PhysRevA.75.012313} {\bibfield  {journal} {\bibinfo
			{journal} {Physical Review A - Atomic, Molecular, and Optical Physics}\
		}\textbf {\bibinfo {volume} {75}},\ \bibinfo {pages} {1} (\bibinfo {year}
		{2007})}\BibitemShut {NoStop}%
	\bibitem [{\citenamefont {Barz}\ \emph {et~al.}(2010)\citenamefont {Barz},
		\citenamefont {Cronenberg}, \citenamefont {Zeilinger},\ and\ \citenamefont
		{Walther}}]{Barz2010}%
	\BibitemOpen
	\bibfield  {author} {\bibinfo {author} {\bibfnamefont {S.}~\bibnamefont
			{Barz}}, \bibinfo {author} {\bibfnamefont {G.}~\bibnamefont {Cronenberg}},
		\bibinfo {author} {\bibfnamefont {A.}~\bibnamefont {Zeilinger}}, \ and\
		\bibinfo {author} {\bibfnamefont {P.}~\bibnamefont {Walther}},\ }\href
	{\doibase 10.1038/nphoton.2010.156} {\bibfield  {journal} {\bibinfo
			{journal} {Nature Photonics}\ }\textbf {\bibinfo {volume} {4}},\ \bibinfo
		{pages} {553} (\bibinfo {year} {2010})}\BibitemShut {NoStop}%
	\bibitem [{\citenamefont {Niu}\ \emph {et~al.}(2009)\citenamefont {Niu},
		\citenamefont {Gong}, \citenamefont {Zou}, \citenamefont {Huang},\ and\
		\citenamefont {Guo}}]{Niu2009}%
	\BibitemOpen
	\bibfield  {author} {\bibinfo {author} {\bibfnamefont {X.~L.}\ \bibnamefont
			{Niu}}, \bibinfo {author} {\bibfnamefont {Y.~X.}\ \bibnamefont {Gong}},
		\bibinfo {author} {\bibfnamefont {X.~B.}\ \bibnamefont {Zou}}, \bibinfo
		{author} {\bibfnamefont {Y.~F.}\ \bibnamefont {Huang}}, \ and\ \bibinfo
		{author} {\bibfnamefont {G.~C.}\ \bibnamefont {Guo}},\ }\href {\doibase
		10.1080/09500340902822341} {\bibfield  {journal} {\bibinfo  {journal}
			{Journal of Modern Optics}\ }\textbf {\bibinfo {volume} {56}},\ \bibinfo
		{pages} {936} (\bibinfo {year} {2009})}\BibitemShut {NoStop}%
	\bibitem [{\citenamefont {H{\"{u}}bel}\ \emph {et~al.}(2010)\citenamefont
		{H{\"{u}}bel}, \citenamefont {Hamel}, \citenamefont {Fedrizzi}, \citenamefont
		{Ramelow}, \citenamefont {Resch},\ and\ \citenamefont
		{Jennewein}}]{Hubel2010}%
	\BibitemOpen
	\bibfield  {author} {\bibinfo {author} {\bibfnamefont {H.}~\bibnamefont
			{H{\"{u}}bel}}, \bibinfo {author} {\bibfnamefont {D.~R.}\ \bibnamefont
			{Hamel}}, \bibinfo {author} {\bibfnamefont {A.}~\bibnamefont {Fedrizzi}},
		\bibinfo {author} {\bibfnamefont {S.}~\bibnamefont {Ramelow}}, \bibinfo
		{author} {\bibfnamefont {K.~J.}\ \bibnamefont {Resch}}, \ and\ \bibinfo
		{author} {\bibfnamefont {T.}~\bibnamefont {Jennewein}},\ }\href {\doibase
		10.1038/nature09175} {\bibfield  {journal} {\bibinfo  {journal} {Nature}\
		}\textbf {\bibinfo {volume} {466}},\ \bibinfo {pages} {601} (\bibinfo {year}
		{2010})}\BibitemShut {NoStop}%
	\bibitem [{\citenamefont {Hamel}\ \emph {et~al.}(2014)\citenamefont {Hamel},
		\citenamefont {Shalm}, \citenamefont {H{\"{u}}bel}, \citenamefont {Miller},
		\citenamefont {Marsili}, \citenamefont {Verma}, \citenamefont {Mirin},
		\citenamefont {Nam}, \citenamefont {Resch},\ and\ \citenamefont
		{Jennewein}}]{Hamel2014}%
	\BibitemOpen
	\bibfield  {author} {\bibinfo {author} {\bibfnamefont {D.~R.}\ \bibnamefont
			{Hamel}}, \bibinfo {author} {\bibfnamefont {L.~K.}\ \bibnamefont {Shalm}},
		\bibinfo {author} {\bibfnamefont {H.}~\bibnamefont {H{\"{u}}bel}}, \bibinfo
		{author} {\bibfnamefont {A.~J.}\ \bibnamefont {Miller}}, \bibinfo {author}
		{\bibfnamefont {F.}~\bibnamefont {Marsili}}, \bibinfo {author} {\bibfnamefont
			{V.~B.}\ \bibnamefont {Verma}}, \bibinfo {author} {\bibfnamefont {R.~P.}\
			\bibnamefont {Mirin}}, \bibinfo {author} {\bibfnamefont {S.~W.}\ \bibnamefont
			{Nam}}, \bibinfo {author} {\bibfnamefont {K.~J.}\ \bibnamefont {Resch}}, \
		and\ \bibinfo {author} {\bibfnamefont {T.}~\bibnamefont {Jennewein}},\ }\href
	{\doibase 10.1038/nphoton.2014.218} {\bibfield  {journal} {\bibinfo
			{journal} {Nature Photonics}\ }\textbf {\bibinfo {volume} {8}},\ \bibinfo
		{pages} {801} (\bibinfo {year} {2014})},\ \Eprint
	{http://arxiv.org/abs/1404.7131} {arXiv:1404.7131} \BibitemShut {NoStop}%
	\bibitem [{\citenamefont {Agne}\ \emph {et~al.}(2017)\citenamefont {Agne},
		\citenamefont {Kauten}, \citenamefont {Jin}, \citenamefont {Meyer-Scott},
		\citenamefont {Salvail}, \citenamefont {Hamel}, \citenamefont {Resch},
		\citenamefont {Weihs},\ and\ \citenamefont {Jennewein}}]{Agne2017}%
	\BibitemOpen
	\bibfield  {author} {\bibinfo {author} {\bibfnamefont {S.}~\bibnamefont
			{Agne}}, \bibinfo {author} {\bibfnamefont {T.}~\bibnamefont {Kauten}},
		\bibinfo {author} {\bibfnamefont {J.}~\bibnamefont {Jin}}, \bibinfo {author}
		{\bibfnamefont {E.}~\bibnamefont {Meyer-Scott}}, \bibinfo {author}
		{\bibfnamefont {J.~Z.}\ \bibnamefont {Salvail}}, \bibinfo {author}
		{\bibfnamefont {D.~R.}\ \bibnamefont {Hamel}}, \bibinfo {author}
		{\bibfnamefont {K.~J.}\ \bibnamefont {Resch}}, \bibinfo {author}
		{\bibfnamefont {G.}~\bibnamefont {Weihs}}, \ and\ \bibinfo {author}
		{\bibfnamefont {T.}~\bibnamefont {Jennewein}},\ }\href@noop {} {\bibfield
		{journal} {\bibinfo  {journal} {Phys. Rev. Lett.}\ }\textbf {\bibinfo
			{volume} {118}},\ \bibinfo {pages} {153602} (\bibinfo {year}
		{2017})}\BibitemShut {NoStop}%
	\bibitem [{\citenamefont {Krapick}\ \emph {et~al.}(2016)\citenamefont
		{Krapick}, \citenamefont {Brecht}, \citenamefont {Herrmann}, \citenamefont
		{Quiring},\ and\ \citenamefont {Silberhorn}}]{Krapick2016}%
	\BibitemOpen
	\bibfield  {author} {\bibinfo {author} {\bibfnamefont {S.}~\bibnamefont
			{Krapick}}, \bibinfo {author} {\bibfnamefont {B.}~\bibnamefont {Brecht}},
		\bibinfo {author} {\bibfnamefont {H.}~\bibnamefont {Herrmann}}, \bibinfo
		{author} {\bibfnamefont {V.}~\bibnamefont {Quiring}}, \ and\ \bibinfo
		{author} {\bibfnamefont {C.}~\bibnamefont {Silberhorn}},\ }\href {\doibase
		10.1364/oe.24.002836} {\bibfield  {journal} {\bibinfo  {journal} {Optics
				Express}\ }\textbf {\bibinfo {volume} {24}},\ \bibinfo {pages} {2836}
		(\bibinfo {year} {2016})}\BibitemShut {NoStop}%
	\bibitem [{\citenamefont {Chekhova}\ \emph {et~al.}(2005)\citenamefont
		{Chekhova}, \citenamefont {Ivanova}, \citenamefont {Berardi},\ and\
		\citenamefont {Garuccio}}]{Chekhova2005}%
	\BibitemOpen
	\bibfield  {author} {\bibinfo {author} {\bibfnamefont {M.~V.}\ \bibnamefont
			{Chekhova}}, \bibinfo {author} {\bibfnamefont {O.~A.}\ \bibnamefont
			{Ivanova}}, \bibinfo {author} {\bibfnamefont {V.}~\bibnamefont {Berardi}}, \
		and\ \bibinfo {author} {\bibfnamefont {A.}~\bibnamefont {Garuccio}},\ }\href
	{\doibase 10.1103/PhysRevA.72.023818} {\bibfield  {journal} {\bibinfo
			{journal} {Physical Review A - Atomic, Molecular, and Optical Physics}\
		}\textbf {\bibinfo {volume} {72}},\ \bibinfo {pages} {1} (\bibinfo {year}
		{2005})}\BibitemShut {NoStop}%
	\bibitem [{\citenamefont {Corona}\ \emph
		{et~al.}(2011{\natexlab{a}})\citenamefont {Corona}, \citenamefont
		{Garay-Palmett},\ and\ \citenamefont {U'Ren}}]{Corona2011}%
	\BibitemOpen
	\bibfield  {author} {\bibinfo {author} {\bibfnamefont {M.}~\bibnamefont
			{Corona}}, \bibinfo {author} {\bibfnamefont {K.}~\bibnamefont
			{Garay-Palmett}}, \ and\ \bibinfo {author} {\bibfnamefont {A.~B.}\
			\bibnamefont {U'Ren}},\ }\href {\doibase 10.1103/PhysRevA.84.033823}
	{\bibfield  {journal} {\bibinfo  {journal} {Physical Review A - Atomic,
				Molecular, and Optical Physics}\ }\textbf {\bibinfo {volume} {84}},\ \bibinfo
		{pages} {033803} (\bibinfo {year} {2011}{\natexlab{a}})}\BibitemShut
	{NoStop}%
	\bibitem [{\citenamefont {Corona}\ \emph
		{et~al.}(2011{\natexlab{b}})\citenamefont {Corona}, \citenamefont
		{Garay-Palmett},\ and\ \citenamefont {U'Ren}}]{Corona2011a}%
	\BibitemOpen
	\bibfield  {author} {\bibinfo {author} {\bibfnamefont {M.}~\bibnamefont
			{Corona}}, \bibinfo {author} {\bibfnamefont {K.}~\bibnamefont
			{Garay-Palmett}}, \ and\ \bibinfo {author} {\bibfnamefont {A.~B.}\
			\bibnamefont {U'Ren}},\ }\href {\doibase 10.1364/OL.36.000190} {\bibfield
		{journal} {\bibinfo  {journal} {Optics Letters}\ }\textbf {\bibinfo {volume}
			{36}},\ \bibinfo {pages} {190} (\bibinfo {year}
		{2011}{\natexlab{b}})}\BibitemShut {NoStop}%
	\bibitem [{\citenamefont {Dot}\ \emph {et~al.}(2012)\citenamefont {Dot},
		\citenamefont {Borne}, \citenamefont {Boulanger}, \citenamefont {Bencheikh},\
		and\ \citenamefont {Levenson}}]{Dot2012}%
	\BibitemOpen
	\bibfield  {author} {\bibinfo {author} {\bibfnamefont {A.}~\bibnamefont
			{Dot}}, \bibinfo {author} {\bibfnamefont {A.}~\bibnamefont {Borne}}, \bibinfo
		{author} {\bibfnamefont {B.}~\bibnamefont {Boulanger}}, \bibinfo {author}
		{\bibfnamefont {K.}~\bibnamefont {Bencheikh}}, \ and\ \bibinfo {author}
		{\bibfnamefont {J.~A.}\ \bibnamefont {Levenson}},\ }\href {\doibase
		10.1103/PhysRevA.85.023809} {\bibfield  {journal} {\bibinfo  {journal}
			{Physical Review A - Atomic, Molecular, and Optical Physics}\ }\textbf
		{\bibinfo {volume} {85}},\ \bibinfo {pages} {023809} (\bibinfo {year}
		{2012})}\BibitemShut {NoStop}%
	\bibitem [{\citenamefont {Zielnicki}\ \emph {et~al.}(2018)\citenamefont
		{Zielnicki}, \citenamefont {Garay-Palmett}, \citenamefont {Cruz-Delgado},
		\citenamefont {Cruz-Ramirez}, \citenamefont {O'Boyle}, \citenamefont {Fang},
		\citenamefont {Lorenz}, \citenamefont {U'Ren},\ and\ \citenamefont
		{Kwiat}}]{Zielnicki2018}%
	\BibitemOpen
	\bibfield  {author} {\bibinfo {author} {\bibfnamefont {K.}~\bibnamefont
			{Zielnicki}}, \bibinfo {author} {\bibfnamefont {K.}~\bibnamefont
			{Garay-Palmett}}, \bibinfo {author} {\bibfnamefont {D.}~\bibnamefont
			{Cruz-Delgado}}, \bibinfo {author} {\bibfnamefont {H.}~\bibnamefont
			{Cruz-Ramirez}}, \bibinfo {author} {\bibfnamefont {M.~F.}\ \bibnamefont
			{O'Boyle}}, \bibinfo {author} {\bibfnamefont {B.}~\bibnamefont {Fang}},
		\bibinfo {author} {\bibfnamefont {V.~O.}\ \bibnamefont {Lorenz}}, \bibinfo
		{author} {\bibfnamefont {A.~B.}\ \bibnamefont {U'Ren}}, \ and\ \bibinfo
		{author} {\bibfnamefont {P.~G.}\ \bibnamefont {Kwiat}},\ }\href {\doibase
		10.1080/09500340.2018.1437228} {\bibfield  {journal} {\bibinfo  {journal}
			{Journal of Modern Optics}\ }\textbf {\bibinfo {volume} {65}},\ \bibinfo
		{pages} {1141} (\bibinfo {year} {2018})}\BibitemShut {NoStop}%
	\bibitem [{\citenamefont {Liscidini}\ and\ \citenamefont
		{Sipe}(2013)}]{Liscidini2013}%
	\BibitemOpen
	\bibfield  {author} {\bibinfo {author} {\bibfnamefont {M.}~\bibnamefont
			{Liscidini}}\ and\ \bibinfo {author} {\bibfnamefont {J.~E.}\ \bibnamefont
			{Sipe}},\ }\href {\doibase 10.1103/PhysRevLett.111.193602} {\bibfield
		{journal} {\bibinfo  {journal} {Physical Review Letters}\ }\textbf {\bibinfo
			{volume} {111}},\ \bibinfo {pages} {193602} (\bibinfo {year}
		{2013})}\BibitemShut {NoStop}%
	\bibitem [{\citenamefont {Fang}\ \emph {et~al.}(2016)\citenamefont {Fang},
		\citenamefont {Liscidini}, \citenamefont {Sipe},\ and\ \citenamefont
		{Lorenz}}]{Fang2016}%
	\BibitemOpen
	\bibfield  {author} {\bibinfo {author} {\bibfnamefont {B.}~\bibnamefont
			{Fang}}, \bibinfo {author} {\bibfnamefont {M.}~\bibnamefont {Liscidini}},
		\bibinfo {author} {\bibfnamefont {J.~E.}\ \bibnamefont {Sipe}}, \ and\
		\bibinfo {author} {\bibfnamefont {V.~O.}\ \bibnamefont {Lorenz}},\ }\href
	{\doibase 10.1007/BF03062537} {\bibfield  {journal} {\bibinfo  {journal}
			{Optics Express}\ }\textbf {\bibinfo {volume} {24}},\ \bibinfo {pages}
		{10013} (\bibinfo {year} {2016})}\BibitemShut {NoStop}%
	\bibitem [{\citenamefont {Fang}\ \emph {et~al.}(2014)\citenamefont {Fang},
		\citenamefont {Cohen}, \citenamefont {Liscidini}, \citenamefont {Sipe},\ and\
		\citenamefont {Lorenz}}]{Fang2014}%
	\BibitemOpen
	\bibfield  {author} {\bibinfo {author} {\bibfnamefont {B.}~\bibnamefont
			{Fang}}, \bibinfo {author} {\bibfnamefont {O.}~\bibnamefont {Cohen}},
		\bibinfo {author} {\bibfnamefont {M.}~\bibnamefont {Liscidini}}, \bibinfo
		{author} {\bibfnamefont {J.~E.}\ \bibnamefont {Sipe}}, \ and\ \bibinfo
		{author} {\bibfnamefont {V.~O.}\ \bibnamefont {Lorenz}},\ }\href {\doibase
		10.1364/OPTICA.1.000281} {\bibfield  {journal} {\bibinfo  {journal} {Optica}\
		}\textbf {\bibinfo {volume} {1}},\ \bibinfo {pages} {281} (\bibinfo {year}
		{2014})}\BibitemShut {NoStop}%
	\bibitem [{\citenamefont {Gravier}\ and\ \citenamefont
		{Boulanger}(2008)}]{Gravier2008}%
	\BibitemOpen
	\bibfield  {author} {\bibinfo {author} {\bibfnamefont {F.}~\bibnamefont
			{Gravier}}\ and\ \bibinfo {author} {\bibfnamefont {B.}~\bibnamefont
			{Boulanger}},\ }\href {\doibase 10.1364/josab.25.000098} {\bibfield
		{journal} {\bibinfo  {journal} {Journal of the Optical Society of America B}\
		}\textbf {\bibinfo {volume} {25}},\ \bibinfo {pages} {98} (\bibinfo {year}
		{2008})}\BibitemShut {NoStop}%
	\bibitem [{\citenamefont {Douady}\ and\ \citenamefont
		{Boulanger}(2004)}]{Douady2004}%
	\BibitemOpen
	\bibfield  {author} {\bibinfo {author} {\bibfnamefont {J.}~\bibnamefont
			{Douady}}\ and\ \bibinfo {author} {\bibfnamefont {B.}~\bibnamefont
			{Boulanger}},\ }\href {\doibase 10.1364/ol.29.002794} {\bibfield  {journal}
		{\bibinfo  {journal} {Optics Letters}\ }\textbf {\bibinfo {volume} {29}},\
		\bibinfo {pages} {2794} (\bibinfo {year} {2004})}\BibitemShut {NoStop}%
	\bibitem [{\citenamefont {Okoth}\ \emph {et~al.}(2019)\citenamefont {Okoth},
		\citenamefont {Cavanna}, \citenamefont {Joly},\ and\ \citenamefont
		{Chekhova}}]{Okoth2018}%
	\BibitemOpen
	\bibfield  {author} {\bibinfo {author} {\bibfnamefont {C.}~\bibnamefont
			{Okoth}}, \bibinfo {author} {\bibfnamefont {A.}~\bibnamefont {Cavanna}},
		\bibinfo {author} {\bibfnamefont {N.~Y.}\ \bibnamefont {Joly}}, \ and\
		\bibinfo {author} {\bibfnamefont {M.~V.}\ \bibnamefont {Chekhova}},\ }\href
	{\doibase 10.1103/PhysRevA.99.043809} {\bibfield  {journal} {\bibinfo
			{journal} {Physical Review A}\ }\textbf {\bibinfo {volume} {99}},\ \bibinfo
		{pages} {1} (\bibinfo {year} {2019})}\BibitemShut {NoStop}%
	\bibitem [{\citenamefont {Drummond}\ and\ \citenamefont
		{Hillery}(1999)}]{Drummond1999}%
	\BibitemOpen
	\bibfield  {author} {\bibinfo {author} {\bibfnamefont {P.~D.}\ \bibnamefont
			{Drummond}}\ and\ \bibinfo {author} {\bibfnamefont {M.}~\bibnamefont
			{Hillery}},\ }\href {\doibase 10.1103/PhysRevA.59.691} {\bibfield  {journal}
		{\bibinfo  {journal} {Physical Review A - Atomic, Molecular, and Optical
				Physics}\ }\textbf {\bibinfo {volume} {59}},\ \bibinfo {pages} {691}
		(\bibinfo {year} {1999})}\BibitemShut {NoStop}%
	\bibitem [{\citenamefont {Drummond}(1990)}]{Drummond1990}%
	\BibitemOpen
	\bibfield  {author} {\bibinfo {author} {\bibfnamefont {P.~D.}\ \bibnamefont
			{Drummond}},\ }\href {\doibase 10.1103/PhysRevA.42.6845} {\bibfield
		{journal} {\bibinfo  {journal} {Physical Review A}\ }\textbf {\bibinfo
			{volume} {42}},\ \bibinfo {pages} {6845} (\bibinfo {year}
		{1990})}\BibitemShut {NoStop}%
	\bibitem [{\citenamefont {Liscidini}\ \emph {et~al.}(2012)\citenamefont
		{Liscidini}, \citenamefont {Helt},\ and\ \citenamefont
		{Sipe}}]{Liscidini2012}%
	\BibitemOpen
	\bibfield  {author} {\bibinfo {author} {\bibfnamefont {M.}~\bibnamefont
			{Liscidini}}, \bibinfo {author} {\bibfnamefont {L.~G.}\ \bibnamefont {Helt}},
		\ and\ \bibinfo {author} {\bibfnamefont {J.~E.}\ \bibnamefont {Sipe}},\
	}\href {\doibase 10.1103/PhysRevA.85.013833} {\bibfield  {journal} {\bibinfo
			{journal} {Physical Review A - Atomic, Molecular, and Optical Physics}\
		}\textbf {\bibinfo {volume} {85}},\ \bibinfo {pages} {013833} (\bibinfo
		{year} {2012})}\BibitemShut {NoStop}%
	\bibitem [{\citenamefont {Yang}\ \emph {et~al.}(2008)\citenamefont {Yang},
		\citenamefont {Liscidini},\ and\ \citenamefont {Sipe}}]{Yang2008}%
	\BibitemOpen
	\bibfield  {author} {\bibinfo {author} {\bibfnamefont {Z.}~\bibnamefont
			{Yang}}, \bibinfo {author} {\bibfnamefont {M.}~\bibnamefont {Liscidini}}, \
		and\ \bibinfo {author} {\bibfnamefont {J.~E.}\ \bibnamefont {Sipe}},\ }\href
	{\doibase 10.1103/PhysRevA.77.033808} {\bibfield  {journal} {\bibinfo
			{journal} {Physical Review A - Atomic, Molecular, and Optical Physics}\ }
		(\bibinfo {year} {2008}),\ 10.1103/PhysRevA.77.033808}\BibitemShut {NoStop}%
	\bibitem [{\citenamefont {Drummond}\ and\ \citenamefont
		{Corney}(1999)}]{Drummond1999a}%
	\BibitemOpen
	\bibfield  {author} {\bibinfo {author} {\bibfnamefont {P.~D.}\ \bibnamefont
			{Drummond}}\ and\ \bibinfo {author} {\bibfnamefont {J.~F.}\ \bibnamefont
			{Corney}},\ }\href {\doibase 10.1364/JOSAB.18.000139} {\bibfield  {journal}
		{\bibinfo  {journal} {Journal of the Optical Society of America B}\ }\textbf
		{\bibinfo {volume} {18}},\ \bibinfo {pages} {7} (\bibinfo {year}
		{1999})}\BibitemShut {NoStop}%
	\bibitem [{\citenamefont {Scully}\ and\ \citenamefont
		{Zubairy}(2008)}]{Scully10}%
	\BibitemOpen
	\bibfield  {author} {\bibinfo {author} {\bibfnamefont {M.~O.}\ \bibnamefont
			{Scully}}\ and\ \bibinfo {author} {\bibfnamefont {M.~S.}\ \bibnamefont
			{Zubairy}},\ }\href@noop {} {\emph {\bibinfo {title} {{Quantum Optics}}}},\
	\bibinfo {edition} {sixth prin}\ ed.\ (\bibinfo  {publisher} {Cambridge
		University Press},\ \bibinfo {address} {United Kingdom},\ \bibinfo {year}
	{2008})\BibitemShut {NoStop}%
	\bibitem [{\citenamefont {Garay-Palmett}\ \emph {et~al.}(2010)\citenamefont
		{Garay-Palmett}, \citenamefont {U'Ren},\ and\ \citenamefont
		{Rangel-Rojo}}]{Garay-Palmett2010}%
	\BibitemOpen
	\bibfield  {author} {\bibinfo {author} {\bibfnamefont {K.}~\bibnamefont
			{Garay-Palmett}}, \bibinfo {author} {\bibfnamefont {A.~B.}\ \bibnamefont
			{U'Ren}}, \ and\ \bibinfo {author} {\bibfnamefont {R.}~\bibnamefont
			{Rangel-Rojo}},\ }\href {\doibase 10.1103/PhysRevA.82.043809} {\bibfield
		{journal} {\bibinfo  {journal} {Physical Review A - Atomic, Molecular, and
				Optical Physics}\ }\textbf {\bibinfo {volume} {82}},\ \bibinfo {pages}
		{043809} (\bibinfo {year} {2010})}\BibitemShut {NoStop}%
	\bibitem [{\citenamefont {Agafonov}\ \emph {et~al.}(2010)\citenamefont
		{Agafonov}, \citenamefont {Chekhova},\ and\ \citenamefont
		{Leuchs}}]{Agafonov2010}%
	\BibitemOpen
	\bibfield  {author} {\bibinfo {author} {\bibfnamefont {I.~N.}\ \bibnamefont
			{Agafonov}}, \bibinfo {author} {\bibfnamefont {M.~V.}\ \bibnamefont
			{Chekhova}}, \ and\ \bibinfo {author} {\bibfnamefont {G.}~\bibnamefont
			{Leuchs}},\ }\href {\doibase 10.1103/PhysRevA.82.011801} {\bibfield
		{journal} {\bibinfo  {journal} {Physical Review A - Atomic, Molecular, and
				Optical Physics}\ } (\bibinfo {year} {2010}),\
		10.1103/PhysRevA.82.011801}\BibitemShut {NoStop}%
\end{thebibliography}
\end{document}